\documentclass{article} 
\usepackage{nips13submit_e,times}
\usepackage{hyperref}
\usepackage{url}

\usepackage{amsmath}
\usepackage{graphicx}
\usepackage{url}
\usepackage{natbib}
\usepackage[noabbrev,capitalise]{cleveref}
\usepackage{lipsum}
\usepackage{natbib}
\usepackage{amssymb}
\usepackage{multicol}
\usepackage{bm}
\usepackage{caption}
\usepackage{wrapfig}
\usepackage{tikz}
\usepackage{float}
\usepackage{subcaption}
\usepackage{multirow}

\newcommand{\NIG}{\mathcal{NW}}
\newcommand{\NW}{\mathcal{NW}}

\usepackage{xcolor}
\definecolor{forestgreen}{rgb}{0.13, 0.55, 0.13}

\usepackage{soul}

\usepackage{setspace}
\title{Generalized Bayesian MARS: Tools for Emulating Stochastic Computer Models}

\author{
Kellin Rumsey$^1$ \\
Statistical Sciences\\
Los Alamos National Laboratories \\
Los Alamos, NM \\
\texttt{knrumsey@lanl.gov} \\
\And
Devin Francom \\
Statistical Sciences\\
Los Alamos National Laboratories \\
Los Alamos, NM  \\
\texttt{dfrancom@lanl.gov} \\
\And
Andy Shen \\
Department of Statistics\\
UC Berkeley \\
Berkeley, CA  \\
}

\nipsfinalcopy 

\begin{document}

\maketitle

\normalsize

\doublespacing
\vspace*{-30pt}

\begin{abstract}
The multivariate adaptive regression spline (MARS) approach of Friedman (1991) and its Bayesian counterpart (Francom et al. 2018) are effective approaches for the emulation of computer models. The traditional assumption of Gaussian errors limits the usefulness of MARS, and many popular alternatives, when dealing with stochastic computer models. We propose a generalized Bayesian MARS (GBMARS) framework which admits the broad class of generalized hyperbolic distributions as the induced likelihood function. This allows us to develop tools for the emulation of stochastic simulators which are parsimonious, scalable, interpretable and require minimal tuning, while providing powerful predictive and uncertainty quantification capabilities. GBMARS is capable of robust regression with t distributions, quantile regression with asymmetric Laplace distributions and a general form of “Normal-Wald” regression in which the shape of the error distribution and the structure of the mean function are learned simultaneously. We demonstrate the effectiveness of GBMARS on various stochastic computer models and we show that it compares favorably to several popular alternatives.
\end{abstract}

\vspace*{0.1cm}\hrule ${}^1$ Corresponding author

\newpage

\standardsize
\doublespacing

\section{Introduction}

Computer models, or simulators, play an important role in many contemporary fields of science. Because computer models can often be expensive to run and sometimes proprietary, practitioners often seek to emulate these models using statistical methods. Stochastic computer models, unlike deterministic models, explicitly seek to model the randomness and uncertainty associated with the process \citep{baker2022analyzing}. Although emulation of deterministic computer models has been extensively studied in the literature, substantially less attention has been given to stochastic models and there are presently many opportunities to create custom tools which can succeed in this area. In particular, the standard assumptions (notably, the Gaussian error distribution) made by many statistical models will often be violated. 

The multivariate adaptive regression spline (MARS) approach of \cite{friedman1991} is a popular tool for nonparametric regression. Modern implementations of the MARS algorithm are fast and powerful methods which have successfully been applied in diverse fields including geoscience, medicine, engineering and finance \citep{zhang2016multivariate, chou2004mining, roy2018estimating, de2011bankruptcy}. In many applications, MARS algorithms (especially modern Bayesian MARS) is comparable and sometimes preferable to other better-known approaches including neural networks, Gaussian process and regression trees \citep{hutchings2023comparing, collins2022bayesian, zhang2016multivariate}. MARS works by modelling the mean of a response as a linear combination of adaptively chosen basis functions. This generally produces models which are flexible and non-parametric while remaining simple to understand and interpret. MARS also performs automatic variable selection and can be recast as an ANOVA decomposition. Finally, MARS often scales quite well with data size compared to alternatives, such as Gaussian process regression \citep{rasmussen2003gaussian}.

In this work, we focus on the Bayesian version of MARS (hereafter BMARS), proposed by \citep{denison1998}, and the improvements discussed by \cite{nott2005} and \cite{francom2018}. Bayesian MARS allows for a comprehensive treatment of uncertainty and is better equipped to combat overfitting, leading to improved prediction in practice.  It has been used extensively as a surrogate for nonlinear computer models \citep{mcclarren2011}, where \cite{chakraborty2013} investigated Gaussian processes with BMARS mean functions, \cite{francom2018} explored the connection to the Sobol decomposition for performing sensitivity analysis, and \cite{francom2019} used it in a large computer experiment with mixed categorical and continuous inputs and spatio-temporal output.  \cite{mallick1999} used BMARS for survival analysis, while \cite{holmes2003classification} used it for classification. In BMARS, a full probability model is specified for the complete set of unknown parameters, and the adaptive selection of basis functions is accomplished using a simplified and effective version of reversible jump Markov chain Monte Carlo \citep{green1995}. Rather than proposing new basis coefficients when the basis changes, the coefficients are integrated out of the posterior making the algorithm relatively easy to implement. This is an important component of the MCMC scheme which we shall revisit shortly.

Modifications to the original BMARS sampling scheme have since been proposed, leading to more efficient exploration of the posterior. For instance, \cite{nott2005} give a more effective proposal distribution for generating basis functions, which is capable of quickly honing in on a particular posterior mode, especially when the number of predictors is large. Other modifications, proposed by \cite{francom2018}, include the use of g-priors and parallel tempering to improve the mixing of the MCMC chain. In each of these improvements to the original sampling scheme, the integration of coefficients out of the posterior has been a critical feature. \cite{holmes2003classification} ensure that they can integrate out coefficients in a classification setting by using a probit link. While \cite{mallick1999} depart from the Gaussian likelihood and thus propose coefficients, doing so comes at great cost both computationally and from the standpoint of inference, as they are no longer confident in their posterior model search. 

In this paper, we propose a generalized BMARS (GBMARS) framework in which the relevant conditional posterior predictive distribution belongs to the broad class of generalized hyperbolic distributions. We accomplish this while maintaining the structure required for efficient MCMC sampling. Although the resulting model requires more training time than traditional BMARS, it retains the excellent scaling properties that make MARS appealing. One special case of the highly-flexible GBMARS framework is Normal-Wald ($\mathcal{NW}$; also called Normal-inverse Gaussian) regression, in which we can simultaneously learn the mean function and (possibly asymmetric and heavy-tailed) error distribution. Other important special cases permit for quantile regression and robust regression (via the $t$ distribution or similar). With these changes, GBMARS can still take advantage of the functional ANOVA decomposition of the GBMARS mean function to understand the influence of the model inputs. The resulting GBMARS then brings the benefits of BMARS, including flexible yet parsimonious mean function modeling, probabilistic uncertainty quantification, and interpretability to the areas of quantile regression, regression with outliers, and regression with unknown but flexible error distribution.  Few similarly scalable and flexible tools exist.

In \cref{sec:BMARS}, we give a brief review of Bayesian MARS, emphasizing the structure required for efficient posterior sampling. In \cref{sec:GBMARS}, we generalize the BMARS framework by introducing a set of latent variables, and we derive the full conditional distributions necessary for posterior sampling. We also discuss a class of prior distributions which, upon marginalization of the latent variables, yields a broad and powerful class of marginal likelihood functions. In \cref{sec:examples}, we analyze various stochastic computer models to emphasize the strong performance of the GBMARS framework in a variety of settings. Concluding remarks are given in \cref{sec:conclusion}.

An implementation of the GBMARS model proposed here can be found at \url{https://github.com/knrumsey/GBASS} and the code used to produce all figures and tables is provided at \url{https://github.com/knrumsey/GBASS-examples}.

\section{Review of Bayesian MARS}
\label{sec:BMARS}

Let $y_i$ denote the response variable corresponding to a vector of $p$ predictor variables $\bm x_i$, for $i=1,\ldots n$. Assume that each predictor variable has been scaled to the unit interval. The response is modeled as
\begin{equation}
\label{eq:bmars}
    \begin{aligned}
        y_i &= f(\bm x_i) + \epsilon_i, \ \epsilon_i \sim N(0, w) \\
        f(\bm x) &= a_0 + \sum_{m=1}^Ma_mB_m(\bm x) \\
        B_m(\bm x) &= \prod_{j=1}^{J_m}\left[s_{j,m}(x_{u_{j,m}} - t_{j,m})\right]_+
    \end{aligned}
\end{equation}
where $s_{j,m} \in \{-1,1\}$ is called a sign, $t_{j,m} \in [0,1]$ is called a knot, $J_m \in \{1,\ldots, J_\text{max}\}$ is the degree of interaction and $u_{j,m} \in \{1,\ldots, p\}$ selects a predictor variable. The function $[\cdot]_+ = \text{max}(0, \cdot)$ is called a Hinge function. We also require that $u_{j,m}$ be distinct for each $j=1,\ldots, J_m$, which implies that $J_{max}$ must be less than or equal to $p$. Finally, $M$ is the number of basis functions, and $\bm a$ is the $M+1$ vector of basis coefficients which includes an intercept. The exhaustive list of parameters we seek to estimate is thus $\bm \theta = \{M, \bm a, w, \bm\theta_B\}$, where $\bm\theta_B = \{ \bm J, \bm s, \bm t, \bm u\}$ represents the basis parameters. $\bm J$ denotes a $M$-vector of interaction degrees, $\bm s$, $\bm t$ and $\bm u$ respectively denote the vector of signs, knots and  variables used, with $\bm s = \{\{s_{j,m}\}_{j=1}^{J_m}\}_{m=1}^M$, and $\bm t$ and $\bm u$ defined similarly.  

First, we consider the prior for the number of basis functions, $M$,
\begin{equation}\label{eq:prior_M}
\begin{aligned}
M|\lambda &\sim \text{Poiss}(\lambda)\times I(M \leq M_\text{max}) \\
\lambda &\sim \text{Gamma}(a_\lambda, b_\lambda).
\end{aligned}
\end{equation}
The truncation value $M_\text{max}$ serves to bound the computational complexity of the algorithm, which is cubic in $M$. The hyperparameters $a_\lambda$ and $b_\lambda$ are the primary tool for addressing overfitting. The usual defaults are $a_\lambda=b_\lambda=10$, although smaller values may be preferred when performing deterministic surrogate modeling (i.e., $\epsilon_i = 0$). 

Next we consider the prior for the regression coefficients $\bm a$. We specify
\begin{align*}
\bm a|w, \tau, \bm\Sigma &\sim N(\bm 0, w\tau\bm \Sigma), \\
    w &\sim \text{InvGamma}(a_w, b_w) \\
    \tau &\sim \text{InvGamma}(a_\tau, b_\tau).
\end{align*}
The Normal prior for $\bm a$ is important from a computational perspective, because the posterior contains the product
\begin{equation}\label{eq:prod_norm}
\begin{aligned}
N(\bm y | \bm B \bm a, w\bm I)N(\bm a | \bm 0, w\tau\bm\Sigma) = 
&\left(w^N\tau^{M+1}\frac{|\bm\Sigma|}{|\bm\Lambda|}\right)^{-1/2} \\
& \times  \ \exp\left\{-\frac{1}{2w}\left(\sum_{i=1}^N\frac{y_i^2}{v_i} - \bm y' \bm B'\bm\Lambda\bm B'\bm y\right) \right\} \\
&\times \ N\left(\bm a \ \bigg| \ \bm\Lambda\bm B'\bm y, \ wc\bm\Lambda\right)
\end{aligned}
\end{equation}
where $\bm\Lambda = \left(\bm B'\bm B + \frac{1}{\tau}\bm\Sigma^{-1}\right)^{-1}$.
The final line of \cref{eq:prod_norm} implies that $\bm a$ can be integrated out of the posterior distribution, avoiding the need for a complicated transdimensional proposal to account for the coefficients.
A ridge prior can be specified by setting $\bm\Sigma = \bm I$, though \cite{francom2020} propose setting $\bm\Sigma = (\bm B'\bm B)^{-1}$ where $\bm B$ is the $n\times (M+1)$ matrix of basis functions which is parameterized by $\bm \theta_B$. They use the default settings $a_\tau = 1/2$, $b_\tau=2/n$ and $a_w = b_w = 0$ which leads to the Zellner-Siow Cauchy prior for $\bm a$. 

Finally, we consider the basis parameters $\bm\theta_B$ which parameterize the basis matrix $\bm B$. First we specify $J_m|M \sim \text{Unif}\{1,\ldots, J_\text{max}\}$ for $m=1,2,\ldots M$, where $J_\text{max}$ is usually at most $3$. For the remaining parameters, we follow \cite{francom2018} and specify 
\begin{equation}
\label{eq:pitheta}
\pi(\bm s_m, \bm t_m, \bm u_m|J_m, M, \bm X) \propto \begin{cases}
\left(\frac{1}{2}\right)^{J_m}\binom{p}{J_m}^{-1}, & b_m \geq b_0 \\
0, & \text{otherwise}
\end{cases}
\end{equation}
where $b_m$ is the number of non-zero values in the basis vector and $b_0$ is a user defined parameter with suggested default $b_0 = 20$, which specifies the minimum number of input points which are allowed to contribute to the local structure of the function. This specification aims to guard against edge instabilities and overfitting.  

At each step of the MCMC algorithm, we will either add a basis function to the model (birth, proposed with probability $p_B$), delete a basis function from the model (death, with probability $p_D$) or modify an existing basis function (mutation, with probability $p_M$). Let $\mathcal M$ denote the current state of the model. A birth step begins by selecting a degree of interaction $J_{M+1}$ from a distribution $I(j|\mathcal M)$ which assigns non-zero mass to the points $\{1,\ldots J_\text{max}\}$. Conditional on this selected value, we sample (i) $s_{j,M+1} \stackrel{\text{iid}}{\sim}\text{Unif}\{-1, 1\}, \ j=1,\ldots J_{M+1}$,
(ii) $t_{j,M+1} \stackrel{\text{iid}}{\sim}\text{Unif}(0, 1), \ j=1,\ldots J_{M+1}$ and (iii) $(u_{1,M+1},\ldots u_{J_{M+1},M+1}) \sim Z(\bm u|\mathcal M)$ where $Z(\bm u|\cdot)$ assigns mass to the set $\{(i_1, \ldots i_{J_{M+1}}) | i_j\in\{1,\ldots p\}, \ i_j\neq i_{j'}\}.$ In other words, sampling from $Z(\bm u|\cdot)$ generates $J_m$ distinct predictor variable indices. The proposal distributions $I(\cdot|\cdot)$ and $Z(\cdot|\cdot)$ are the primary contribution of \cite{nott2005}, and lead to more efficient posterior exploration. In particular, we specify
$$I(j|\mathcal M) \propto w_1 + \sum_{m=1}^MI(J_m= j),$$
so that interaction degrees are proposed in a manner which is (except for $w_1$) proportional to the number of times each interaction degree has already appeared in the model. As $w_1\rightarrow\infty$, the proposal is uniform over the set $\{1,\ldots J_\text{max}\}$, which is consistent with the original BMARS sampling scheme. The distribution $Z(\cdot|\cdot)$ is similar, but necessarily more complicated. If we define
$$z(u|\mathcal M) = w_2 + \sum_{m=1}^M\sum_{j=1}^{J_m}I(u_{j,m}=u),$$
then $Z(\bm u|\mathcal M)$ proposes $J_{M+1}$ distinct values from the set $\{1,\ldots p\}$ by sampling without replacement and with sampling weights $z(u|\mathcal M)$. This means that variables which have frequently been incorporated into the model are more likely to be proposed. More concisely, the distribution $Z(\bm u|\cdot)$ can be recognized as Wallenius' noncentral hypergeometric distribution \citep{fog2008calculation}. Small values of the tuning parameters are likely to converge to a posterior mode more quickly, although choosing too small a value can cause the sampler to get stuck in a suboptimal region of the posterior. \cite{francom2020} recommend setting $w_1 = w_2 = 5$.

The death and mutation steps are considerably easier to describe. In the former, a basis function is selected uniformly at random and deleted from the current model. In mutation, the degree of interaction and the variables indices are held constant, while the knots and signs are modified. In all three cases, we let the $\star$ subscript denote the modified state of the model. These changes will be accepted as the new state with log-probability
\begin{equation}
\label{eq:accept_gen}
\begin{aligned}
\log\alpha_X &= -\frac{1}{2}\Bigg[\log|\bm\Sigma_\star| - \log|\bm\Sigma| - \log|\bm\Lambda_\star| + \log|\bm\Lambda| \\
& - \frac{1}{w}\left( \bm z'\left[\bm B_\star'\bm\Lambda_\star \bm B_\star - \bm B' \bm\Lambda \bm B\right] \bm z\right)\Bigg]\\
 & + \log S_X
 \end{aligned}
\end{equation}
where $X \in \{B, D, M\}$ indicates the move type and $S_X$ is given for the more general case in \cref{eq:accept_genS}. See \cite{denison1998}, \cite{nott2005}, and the supplementary material to \cite{francom2019} for additional examples of deriving the RJMCMC acceptance probabilities for BMARS.  Once the proposed modifications have been accepted or rejected, the remaining parameters $(\bm a, w, \lambda, \tau)$ can be sampled efficiently from their full conditional distributions. The full conditional distributions will be given in the more general case in Section 3. 

\section{Generalized Bayesian MARS}
\label{sec:GBMARS}

In this section, we will present a generalized BMARS framework which permits a wide range of marginal likelihoods. In order to maintain a computationally efficient MCMC sampler, we employ a useful trick and specify the model as a Normal mean-variance mixture representation \citep{barndorff1982normal}. The full model, for $i=1,\ldots, n$, can be stated as
\begin{equation}
\label{eq:gbmars_error}
\begin{aligned}
    y_i &= a_0 + \sum_{m=1}^Ma_m\prod_{j=1}^{J_m}[s_{j,m}(x_{u_{j,m}} - t_{j,m})]_+ + \epsilon_i \\ 
    \epsilon_i &= \sqrt{w}\left(\beta v_i  + \sqrt{v_i} \zeta_i\right), \quad \zeta_i \stackrel{\text{iid}}{\sim} N(0, c) 
\end{aligned}  
\end{equation}
with priors
\begin{equation}
\label{eq:gbmars_priors}
\begin{aligned}
    \bm a | w, \tau &\sim N(\bm 0, w\tau\bm\Sigma) \hspace{1cm}
    \tau \sim \text{InvGamma}(a_\tau, b_\tau) \\
    M|\lambda &\sim \text{Poiss}(\lambda) \hspace{1.45cm}
    \lambda \sim \text{Gamma}(a_\lambda, b_\lambda) \\
    \bm\theta_B = \{ \bm J, \bm s, \bm t, \bm u\} &\sim \pi_\theta \hspace{2.3cm}
    \beta \sim N(m_\beta, s_\beta^2) \\
    w &\sim \pi_w  \hspace{2.2cm}
    v_i \stackrel{\text{iid}}{\sim} \pi_v.
\end{aligned}  
\end{equation}

We refer to $w$ and $v_i$ respectively as the global and local variance factors, because $w$ describes the variance of all responses while $v_i$ influences the likelihood of just the $i^{th}$ response. The priors for $v_i$ and $w$ will be discussed in the next subsection. The parameter $\beta$ allows for asymmetry in the marginal likelihood function and is referred to as a skewness parameter, and $c$ is a fixed constant (often set to unity) which is included for convenience, as will become clear in later sections. We assign to $\bm\theta_B$ the same prior as in \cref{eq:pitheta}.

By defining $z_i = y_i - \beta v_i\sqrt w$, the generalized model can be expressed as $z_i = f(\bm x_i) + \epsilon_i, \ \epsilon_i \sim N(0, cwv_i)$, which looks very similar to the BMARS model of \cref{eq:bmars}. In matrix form, we can write the GBMARS model as $\bm z|\cdot \sim N\left(\bm B \bm a, cw\bm V\right)$ where $\bm V$ is a diagonal matrix whose $(ii)^{th}$ component is $v_i$. We will sample from this model using an algorithm which is very similar to the one described in \cref{sec:BMARS}, and any departures will be discussed. This latent variable structure is useful because the coefficients can still be integrated out of the posterior, although the marginal distribution for $\bm a$ is a bit more complicated than before (see the supplemental materials (SM1) for details). The acceptance probabilities for the birth, death and mutation steps also look slightly different, with
\begin{equation}
\label{eq:accept_gen2}
\begin{aligned}
\log\alpha_X &= -\frac{1}{2}\Bigg[\log|\bm\Sigma_\star| - \log|\bm\Sigma| - \log|\bm\Lambda_\star| + \log|\bm\Lambda| \\
& \quad - \frac{1}{wc}\left( \bm z'\bm V^{-1}\left[\bm B_\star'\bm\Lambda_\star \bm B_\star - \bm B' \bm\Lambda \bm B\right]\bm V^{-1} \bm z\right)\Bigg]  + \log S_X,
 \end{aligned}
\end{equation}
Where  $\bm\Lambda = \left(\bm B'\bm V^{-1}\bm B + \frac{c}{\tau}\bm\Sigma^{-1}\right)^{-1}$ and each $S_X$ (for $X \in \{B, D, M\}$) can be computed as
\begin{equation}
\label{eq:accept_genS}
\begin{aligned}
S_B &= \frac{c^{1/2}p_D\lambda 1(b_m \geq b_0)}{\tau^{1/2}p_B(M+1)J_{max}\binom{p}{J_{M+1}}I(J_{M+1}|\mathcal M)z(\bm u_{M+1}|\mathcal M)} \\[1.1ex]
S_D &= \frac{\tau^{1/2}p_BMJ_{max}\binom{p}{J_{m_\star}}I(J_{m_\star}|\mathcal M_\star)z(\bm u_{m_\star}|\mathcal M_\star)}{c^{1/2}p_D\lambda} \\[1.1ex]
S_M &= 1.
 \end{aligned}
\end{equation}

We also note that the choice $\bm\Sigma = (\bm B'\bm B)^{-1}$ no longer has the same computational appeal as it previously did. It may seem tempting to specify a modified g-prior by setting $\bm\Sigma = (\bm B'\bm V^{-1}\bm B)^{-1}$ as explored by \cite{alhamzawi2015}, but this significantly complicates the full conditional distribution of the local variance components and will have a substantial impact on the efficiency of the MCMC. Thus, we advocate for the simple choice $\bm\Sigma = \bm I$ in the GBMARS framework, which corresponds to a ridge prior. 

\subsection{Full Conditional Distributions}

We are now ready to complete the specification of the posterior sampler by stating the full conditional distribution for each of the parameters in the model which are updated via a Gibbs step. The priors for $w$ and $v_i$ are left in a general form right now and will be discussed shortly. The proposal and prior distributions for $\bm\theta_B$ are the same as in \cref{sec:BMARS}. For notational convenience, we define $\hat{\bm y} = \bm B\bm a$ and $\bm r = \hat{\bm y} - \bm y$. 
\begin{equation}\label{eq:gibbs}
\begin{aligned}
\bm a | \cdot &\sim  N\left( \bm\Lambda\bm B'\bm V^{-1}\bm z, \ wc\bm\Lambda\right) \\[1.4ex]
\lambda | \cdot &\stackrel{\text{aprx}}{\sim} \text{Gamma}\left(a_\lambda + M, b_\lambda + 1 \right) \\[1.4ex]
\tau | \cdot &\sim \text{Inv-Gamma}\left(a_\tau + \frac{M+1}{2}, b_\tau + \frac{\bm a' \bm\Sigma^{-1}\bm a}{2w}\right) \\[1.4ex]
\beta|\cdot &\sim N\left(\frac{s_\beta^2 \sum r_i/\sqrt w + m_\beta c}{s_\beta^2\sum v_i + c}, \ \frac{s_\beta^2c}{s_\beta^2\sum v_i + c}\right) \\[1.4ex]
\pi(w|\cdot) &\propto w^{-(N+M+1)/2}\exp\left\{-\frac{1}{2}\left(\frac{1}{c}\sum_{i=1}^N\frac{r_i^2}{v_i} + \frac{1}{\tau}\bm a'\bm\Sigma^{-1}\bm a\right)\frac{1}{w}\right\}\\ 
&\hspace*{2cm}\times\exp\left\{\frac{\beta }{c}\sum_{i=1}^Nr_i \frac{1}{\sqrt w}\right\}\pi_w(w) \\[1.4ex]
\pi(v_i|\cdot) &\propto v_i^{-1/2}\exp\left\{-\frac{1}{2}\left(\frac{\beta ^2}{c}v_i + \frac{r_i^2}{wc}\frac{1}{v_i}\right)\right\} \pi_v(v_i), \ i=1,\ldots n
\end{aligned}
\end{equation}
where the full conditional for $\lambda$ is approximate because of the truncation of the Poisson prior for $M$.  This formulation for $\lambda$ is used in other BMARS implementations, and the approximation is accurate as long as the number of basis functions does not get close to the upper bound.  Typically, if $M$ is approaching the upper bound, we would either increase the upper bound or change the model specification elsewhere to get a more parsimonious model.  

\subsection{Prior for the Local Variance Factors}
\label{sec:priorlocal}
By marginalizing over the prior distribution for the latent $v_i$, a broad class of useful marginal likelihoods can be obtained. The choice of a generalized inverse Gaussian (GIG) distribution for $\pi_v$ yields the powerful class of generalized hyperbolic distributions as the marginal likelihood. The generalized hyperbolic distribution contains many interesting special cases and will be discussed in \cref{sec:marginal}. Additionally, the GIG prior is conjugate for each of the $v_i$ and fast random generators exist, making it an excellent candidate in the present setting \citep{devroye2014random, hormann2014generating}.

 The density of the GIG distribution is given by 
\begin{equation}
\label{eq:GIG}
\text{GIG}(x | p,a,b)= \frac{(a/b)^{\frac{p}{2}}}{2K_p\left(\sqrt{ab}\right)} x^{p-1}\exp \left(-\frac{1}{2}\left(ax + b\frac{1}{x}\right)\right), \ p \in \mathbb R, a, b \geq 0,
\end{equation}
which includes the Gamma ($b=0$, $p > 0$), Inverse Gamma ($a=0$, $p<0$) and Wald (also called Inverse-Gaussian $p=-1/2, a>0, b>0$) distributions as special cases \citep{jorgensen2012}.  Many generators have been proposed for sampling from the GIG distribution, but we prefer the rejection sampling approach of \cite{hormann2014generating}, which is uniformly bounded and specifically designed for the case of varying parameters (e.g., Gibbs sampling). An efficient implementation using C can be found in the \textsf{R} package \textsf{GIGrvg} \citep{leydold2017package}. If we specify the prior $v_i \stackrel{\text{iid}}{\sim} \text{GIG}(p_v, a_v, b_v)$, then the full conditional distribution of each $v_i$ can be written as 
\begin{equation}
    v_i|\cdot \sim \text{GIG}\left(p_v - \frac{1}{2}, a_v + \frac{\beta^2}{c}, b_v + \frac{r_i^2}{wc}\right).
\end{equation}

The generalized beta prime (GBP) distribution is an intriguing alternative to the GIG prior, and has some interesting connections with the Horseshoe prior \citep{carvalho2009}. We have had some success with this approach in certain problems and more details are provided in the supplemental materials (SM2).

\subsection{Prior for the Global Variance Factor}
When the asymmetry parameter $\beta$ is fixed at zero ($m_\beta = s_\beta = 0$) and the marginal likelihood is symmetric, the GIG prior is also conjugate for the global variance factor $w$. That is, when $\beta = 0$ and when we specify the prior $w \sim \text{GIG}(p_w, a_w, b_w)$, the full conditional posterior of $w$ becomes 
\begin{equation}
    w|\cdot \sim GIG\left(p_w - \frac{N+M+1}{2}, a_w, b_w + \frac{1}{c}\sum_{i=1}^N\frac{r_i^2}{v_i} + \frac{1}{\tau}\bm a'\bm\Sigma^{-1}\bm a\right).
\end{equation}
In the more general case where $\beta \neq 0$, the full conditional of $w$ is somewhat more complicated. It can be demonstrated, however, that the density $\pi(w|\cdot)$ is log-concave whenever $n + M + p_w > 2$, and we use the rejection sampling approach of \citep{devroye2014random} to generate samples efficiently from the conditional posterior. In particular, if $p_w < 0, a_w = 0$ and $b_w > 0$, then $w$ can be sampled as 
\begin{equation}
    w^{-1/2} \sim \text{MHN}\left(N+M-2p_w+1, \frac{1}{2}\left(b_w + \frac{1}{c}\sum_{i=1}^N\frac{r_i^2}{v_i} + \frac{1}{\tau}\bm a^\prime\bm\Sigma^{-1}\bm a\right), \frac{\beta}{c}\sum_{i=1}^Nr_i\right) 
\end{equation}
for arbitrary $\beta$, where MHN$(\cdot, \cdot, \cdot)$ denotes the modified half Normal distribution as described by \cite{sun2021modified}. This distribution is always log-concave and can be sampled from efficiently. 

The GIG prior encompasses a wide variety of useful scale priors and is therefore a reasonable choice. As with the local variance factors, however, the GBP prior is an interesting alternative and is discussed in the supplemental materials.

\subsection{The Marginal Likelihood}
\label{sec:marginal}
The GBMARS framework described above incorporates a large and exciting class of induced likelihood functions. Conditional on all of the parameters, the observed response variable is a Gaussian random variable, e.g. $y_i|\cdot \sim N(A + Bv_i, Cv_i)$ for suitable constants $A$, $B$ and $C$. By integrating over the prior for the nuisance variable $v$, the induced marginal distribution for $y$ is given by $\int_0^\infty N(y|A+Bv,Cv)\text{GIG}(v|p_v, a_v, b_v)dv$. The resulting distribution, which was first studied in \cite{barndorff1977exponentially}, is known as the generalized hyperbolic (GH) distribution. The GH distribution is a flexible $5$ parameter distribution which captures a wide range of behaviors optionally allowing for skewness and heavy tails. Notable subclasses include the Normal-Wald ($\NIG$), hyperbolic, variance-gamma, asymmetric Laplace, Gaussian, $t$, double exponential (or Laplace) and logistic distributions. \Cref{tab:marginal} summarizes some of the possible induced likelihoods as a function of the relevant parameters. 

We note that the mean and variance of $y_i$ can be written as
\begin{equation}
    \begin{aligned}
        E(y_i) &= E\left(E(y_i|v_i)\right) \\
        &= (\bm B \bm a)_i + \sqrt{w}\beta E(v_i) \\[1.5ex]
        \text{Var}(y_i) &= E\left( \text{Var}(y_i|v_i)\right) +  \text{Var}\left(E(y_i|v_i)\right) \\
        &= w\left(c E(v_i) + \beta^2 \text{Var}(v_i)\right)
    \end{aligned}
\end{equation}
where $E(v_i)$ and $\text{Var}(v_i)$ are the mean and variance of $v_i$ under the specified GIG prior, which can be readily computed using modified Bessel's functions. This is important, especially when $\beta \neq 0$, because the predictions based on $\bm B\bm a$ may be biased. Thus posterior inference should be conducted with the full distribution of $\bm \epsilon$ in mind, for instance by constructing posterior predictive intervals, or by adjusting the posterior predictions accordingly. 

\Cref{tab:marginal} illustrates that there are a wide variety of likelihoods that are worth exploring, but we will limit our discussion to just three cases of interest. First, we consider the case where the output of the computer model is $\mu(\bm x) + \epsilon_i$ where the error terms, $\epsilon_i$, are iid but possibly non-Gaussian. In this setting, we propose fitting the surrogate under a $\NIG$ likelihood distribution. It is a challenging problem, but in many cases we can recover the mean function $\mu(x)$ as well as an excellent approximation to the error distribution. This will lead to well-calibrated posterior prediction intervals which are much more precise than emulators which make the standard iid Gaussian error assumption. Secondly, we will demonstrate that the GBMARS framework provides a state-of-the-art method for non-linear quantile regression. Quantile regression can readily characterize heteroskedasticity, skewness and heavy tails of the error distribution and is a popular approach for treating stochastic computer models. Quantile regression with GBMARS is often superior to popular competitors, such as quantile kriging \citep{plumlee2014building}, especially when there are few replications in the training data. Finally, we consider regression under a variety of heavy-tailed likelihoods, such as the $t$ distribution. For computer models which display heavy tailed responses, or potential outliers, the mean function can be recovered more robustly by using such an approach.

\begin{table}[t]
  \centering 
  \caption{A selection of marginal likelihoods available in the GBMARS framework after marginalizing over $v$. Scale parameter $w=1$ for simplicity.}
\begin{tabular}{@{\extracolsep{15pt}} ccccl} 
\\[-1.8ex]\hline 
\hline \\[-1.8ex] 
$\beta$ & $p_v$ & $a_v$ & $b_v$ & Marginal Likelihood \\ 
\hline \\[-1.8ex] 
$\cdot$ & $1$ & $a$ & $b$ & Hyperbolic \\[1.2ex]
$\cdot$ & $\frac{-1}{2}$ & $\gamma^2$ & $\delta^2$ & $\NIG$ \\[1.2ex]
$\cdot$ & $\frac{3}{\kappa}$ & $\frac{6}{\kappa}$ & $0$ & Variance-Gamma$(\beta, 1, \kappa)$ \\[1.2ex]
$\cdot$ & $1$ & $\frac{2}{\lambda}$ & $0$ & Asymmetric-Laplace($\lambda$, $\beta$) \\[1.2ex]
$0$ & $\infty$ & $\infty$ & 0 & Gaussian \\[1.2ex]
$0$ & $\frac{-\nu}{2}$ & 0 & $\frac{\nu}{2}$ & $t(\nu)$ \\[1.2ex]
$0$ & $1$ & $\frac{2}{\lambda}$ & $0$ & Double Exponential \\[1.2ex]
$0$ & $\frac{5}{2}$ &  $\frac{15}{\pi^2}$ & 0 & (Approximately) Logistic \\[1.2ex]
\hline \\[-1.8ex] 
\end{tabular} 
\label{tab:marginal}
\end{table} 

\subsubsection{Normal Wald Regression}
Regression using a Normal-Wald ($\NIG$; also called Normal-inverse Gaussian) likelihood allows for discovery of various tail behaviors, encompassing many single-peaked iid error distributions, without requiring significant \textit{a priori} knowledge. Popularized by \cite{barndorff1997normal} in the field of mathematical finance, the $\NIG$ is flexible enough to handle a wide range of distributions, including those with heavy tails or skewness. The $\NIG$ density function is given by
\begin{equation}
    \label{eq:NIG}
    \NIG(x | \alpha, \beta, \delta, \mu) = \frac{\alpha\delta K_1\left(\alpha\sqrt{\delta^2 + (x-\mu)^2}\right)}{\pi\sqrt{\delta^2+(x-\mu)^2}}\exp\left(\delta\sqrt{\alpha^2-\beta^2}+\beta(x-\mu)\right),
\end{equation}
where $K_1(\cdot)$ is the modified Bessel function of the second kind, $\alpha$ is a tail heaviness parameter, $\beta$ is an asymmetry parameter, $\delta$ is a scale parameter and $\mu$ is a location parameter. As shown in \cref{tab:marginal}, the $\NIG$ distribution is a member of the GBMARS framework, by setting
$v_i \stackrel{\text{iid}}{\sim} \text{GIG}\left(-1/2, \gamma^2, \delta^2 \right),$
which corresponds to 
\begin{equation}
\epsilon_i|\gamma,\beta,w \sim \NIG\left(\sqrt{\frac{\beta^2+\gamma^2}{w}}, \frac{\beta}{\sqrt w},  \sqrt w\delta,0\right).    
\end{equation}
Since $\delta$ and $w$ are both scale parameters, we recommend setting $\delta = 1$ to maintain identifiability. For $\beta$, we recommend the default prior settings $m_\beta = 0$ and $s_\beta = 100$, although relevant prior information may be useful here. The hyperparameter $\gamma$, which controls the tail behavior of the $\NIG$ distribution, can either be fixed at a reasonable value or can be assigned a Normal prior $\gamma \sim N(m_\gamma, s_\gamma)$ leading to the conjugate update 
\begin{equation}
    \gamma|\cdot \sim N\left(\frac{s_\gamma^2 n + m_\gamma}{s_\gamma^2\sum_{i=1}^nv_i + 1}, \frac{s_\gamma^2}{s_\gamma^2\sum_{i=1}^nv_i + 1}\right).
\end{equation}

Selecting values for the hyperparameters $m_\gamma$ and $s_\gamma$ is a non-trivial task, and can be crucial to the success of $\NIG$ regression. To clarify the problem, we find that the $\NIG$ shape triangle leads to some useful insights \citep{rydberg1997normal}. The $\NIG$ shape triangle refers to the domain of the transformed parameters $(\chi, \xi)$, where $\chi$ and $\xi$ are called asymmetry and steepness parameters respectively. These interpretable parameters are analogous to skewness and kurtosis, but are invariant under location and scale transformations. These parameters are defined as
\begin{equation}
    \label{eq:NIG_triangle}
    \xi = (1+|\gamma|)^{-1/2} \quad\quad\text{and}\quad\quad \chi = \frac{\beta\xi}{\alpha}.
\end{equation}
The domain of variation for $(\chi, \xi)$ is $0 \leq |\chi| < \xi < 1$, and includes both the Gaussian ($\chi=0, \xi \rightarrow 0$) and the Cauchy ($\chi=0, \xi\rightarrow 1$) distribution as limiting cases. Since small values of $\xi$ represent Gaussian tails and large values of $\xi$ represent Cauchy tails, it is fairly intuitive to specify a Beta prior for $\xi$. Since $|\gamma| = \frac{1}{\xi^2} - 1$, we can match quantiles of the left hand side (folded Normal) and right hand side (transformation of Beta) to obtain reasonable estimates of $m_\gamma$ and $s_\gamma$. For example, the choice $\xi \sim \text{Beta}(1, 5)$ suggests a moderate preference towards Gaussian tails and leads to hyperparameters $m_\gamma \approx 90$ and $s_\gamma \approx 25$, which we will suggest as a default. More details are given in SM3 of the supplemental materials.

\subsubsection{Quantile Regression}
\label{sec:quantileregression}
Bayesian quantile regression is often performed by obtaining the posterior under an asymmetric Laplace likelihood \citep{yu2001bayesian, kotz2012}. This is analogous to the frequentist approach of minimizing the so-called ``pinball-loss" $\sum_{i=1}^nx_i(q - I(x < 0))$, which generalizes absolute loss as in regression for the median. This is readily accomplished within the GBMARS framework by setting
\begin{equation}
    p_v = 1, \quad\quad a_v = 2, \quad\quad b_v = 0, \quad\quad c = \frac{2}{q(1-q)}, \quad\quad m_\beta = \frac{1-2q}{q(1-q)}, \quad\quad s_\beta = 0.
\end{equation}

\begin{figure}[t]
    \centering
    \includegraphics[width=0.85\textwidth]{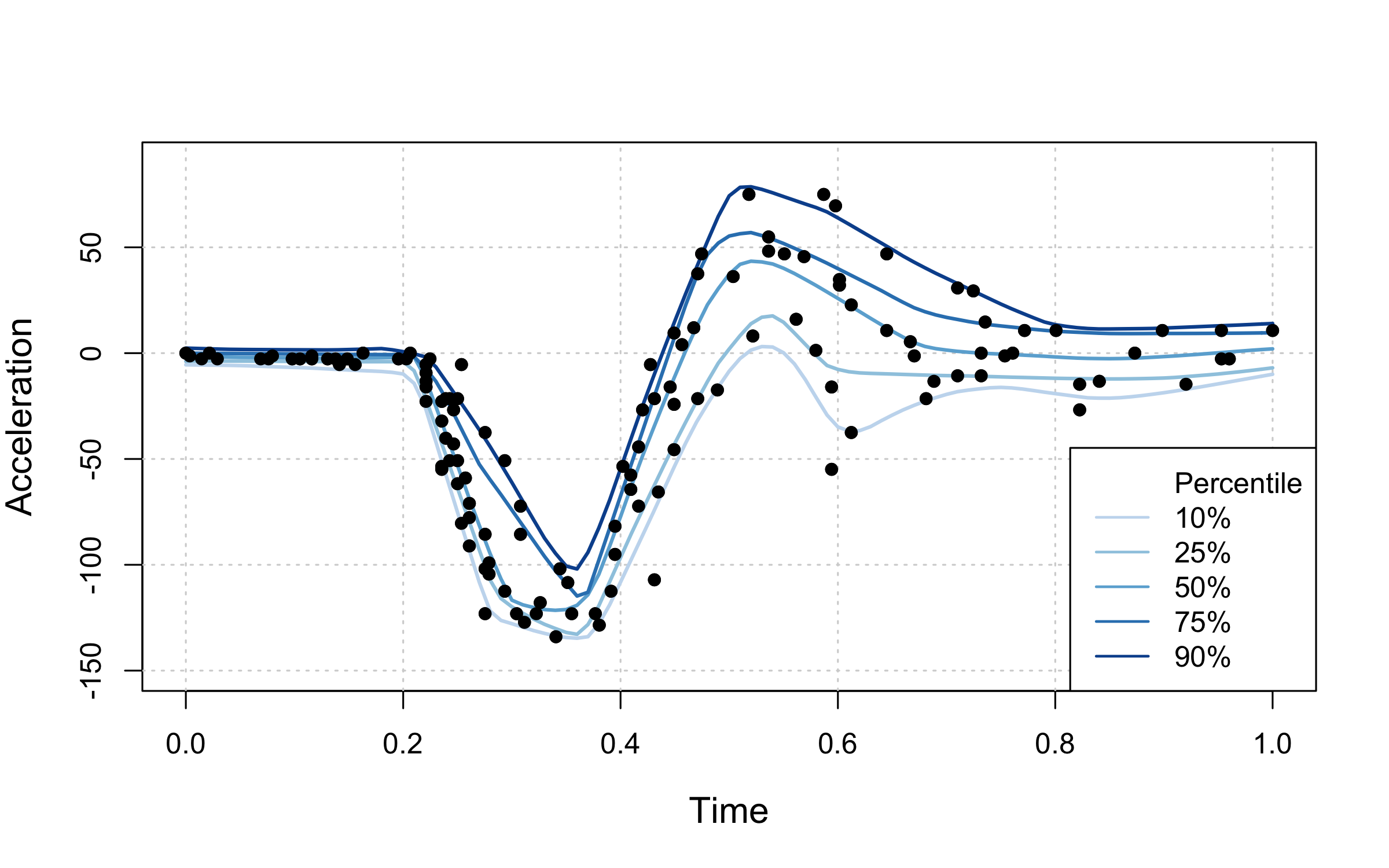}
    \caption{Quantile regression with GBMARS on the motorcycle dataset. Note that GBMARS does not require replications in the predictor variables.}
    \label{fig:mcycle}
\end{figure}

We will demonstrate in \Cref{sec:examples} that GBMARS provides a powerful and accurate tool for quantile regression. A highly desirable feature of the method is that, unlike many other approaches, the computer model does not need to be evaluated (replicated) multiple times for the same input value. We will formally evaluate this approach in \cref{sec:examples}, but \Cref{fig:mcycle} presents a visual demonstration of quantile regression with GBMARS for the well-known {\it motorcycle} dataset of \cite{silverman1985}. 

As is common with many quantile regression techniques, GBMARS is prone to quantile crossing \citep{he1997quantile}, although we find it seldom occurs in practice. We note that the tendency for quantile crossing to occur can be reduced by using replications in the training data. Selecting small values of the hyperparameters $a_\lambda$ and $b_\lambda$, to allow for a more flexible fit via more basis functions, also increases the chances of non-crossing quantiles. Thus for quantile regression with GBMARS, we suggest the less informative default values of $a_\lambda = b_\lambda = 0.01$.

\subsubsection{Heavy Tailed Regression}
Stochastic computer models and real world data are often plagued by heavy tails and/or extreme outputs (outliers). In either case, it is well known that regression under a Gaussian likelihood can by hyper-sensitive to the heavy tails or corrupted data, leading to poor performance \citep{li1985robust, andersen2008modern}. Robust regression under Student's $t$ likelihood is probably the most popular choice, but the logistic and variance-gamma distributions provide interesting alternatives. Regression for the median, which is a special case of \cref{sec:quantileregression} using a double exponential (symmetric Laplace) likelihood, also exhibits improved robustness to non-Gaussian errors. 

In this context, we generally restrict ourselves to symmetric likelihoods (by setting $m_\beta = s_\beta = 0$) for simplicity. The $t(\nu)$, logistic, variance-gamma and double exponential distributions can all be readily obtained depending on the choice of hyperparameters $p_v, a_v$ and $b_v$. See \cref{tab:marginal} for details. 

\section{Examples}
\label{sec:examples}

In this section, we demonstrate the utility of GBMARS using a variety of examples. First, we compare GBMARS under the $\NW$ and $t$ likelihoods to some other popular emulators for a simulator which we make stochastic with non-Gaussian errors. In this setting, we also conduct a scaling study to show that GBMARS scales favorably with the size of the training data. Next, we use a physical agent-based model to demonstrate the effectiveness of GBMARS as a tool for quantile regression. Finally, we leverage the connection between MARS models and the Sobol decomposition and perform a novel sensitivity study for a simple stochastic compartmental disease model. The R code used to generate all of these examples is available at \url{https://github.com/knrumsey/GBASS-Examples}. For every GBMARS emulator fit in this section, we use the default settings described in \cref{sec:GBMARS}. The MCMC sampler is run for $10,000$ iterations, discarding the first $9,000$. 

Additional examples, including (i) analysis of regression in the presence of outliers, (ii) analysis of quantile regression in high dimensions and (iii) $\mathcal NW$ regression on several real world data sets can be found in sections SM4-SM6 of the supplemental material.  

\subsection{The Piston Function}
In this example we will demonstrate that (i) $\NIG$-regression with GBMARS is capable of simultaneously discovering both the mean and error distribution of a stochastic computer model with iid (non-Gaussian) errors and (ii) that this often leads to superior emulation, especially in terms of uncertainty quantification, compared to other popular alternatives which assume Gaussian errors. We also use this framework to perform a scaling study, in which we show that GBMARS generally scales nicely with $n$, although the overhead is larger than alternatives such as Gaussian BMARS and BART.  

The piston function simulates the cycle time, in minutes, of a piston using $7$ inputs \citep{zacks1998modern, ben2007modeling}. The inputs and their usual ranges are given as follows: $M\sim \text{Unif}(30, 60)$ (piston weight (kg)), $S \sim \text{Unif}(0.005, 0.02)$ (piston surface area ($m^2$)), $V_0\sim \text{Unif}(0.002, 0.01)$ (initial gas volume ($m^3$)), $k\sim \text{Unif}(1000, 5000)$ (spring coefficient (N/m), $P_0 \sim \text{Unif}(90000, 110000)$ (Atmospheric pressure (N/$m^2$), $T_a\sim \text{Unif}(290, 296)$ (ambient temperature (K)) and $T_0\sim \text{Unif}(340, 360)$ (filling gas temperature (K)). The piston function is defined as 
\begin{equation}
    \label{eq:piston}
\begin{aligned}
    f(\bm x) &= 120\pi\sqrt{\frac{M}{k + S^2\frac{P_0V_0T_a}{T_0V^2}}} \\
    V &= \frac{S}{2k}\left(\sqrt{A^2+4k\frac{P_0V_0T_a}{T_0}}-A\right)\text{ and } A = P_0S+19.62M-\frac{kV_0}{S}
\end{aligned}
\end{equation}

\subsubsection{Recovering the Error Distribution}
We will assume that a stochastic version of the piston simulator, which takes high-order interactions and dynamic conditions into account, may reasonably return outputs of the form
$y_i = f(\bm x_i) + \epsilon_i$
where $\epsilon_i$ is a highly non-Gaussian error term. To conduct a through examination, we generate training data for $n=1,000$ locations using a maximin Latin hypercube design and we generate the iid $\epsilon_i$ using an asymmetric Laplace error distribution with mean $0$, standard deviation $0.0812$ and skewness $-1.8$ (the standard deviation was chosen so that the signal to noise ratio is approximately $3$). 

\begin{figure}
\centering
\begin{subfigure}{.5\textwidth}
  \centering
  \includegraphics[width=.95\linewidth]{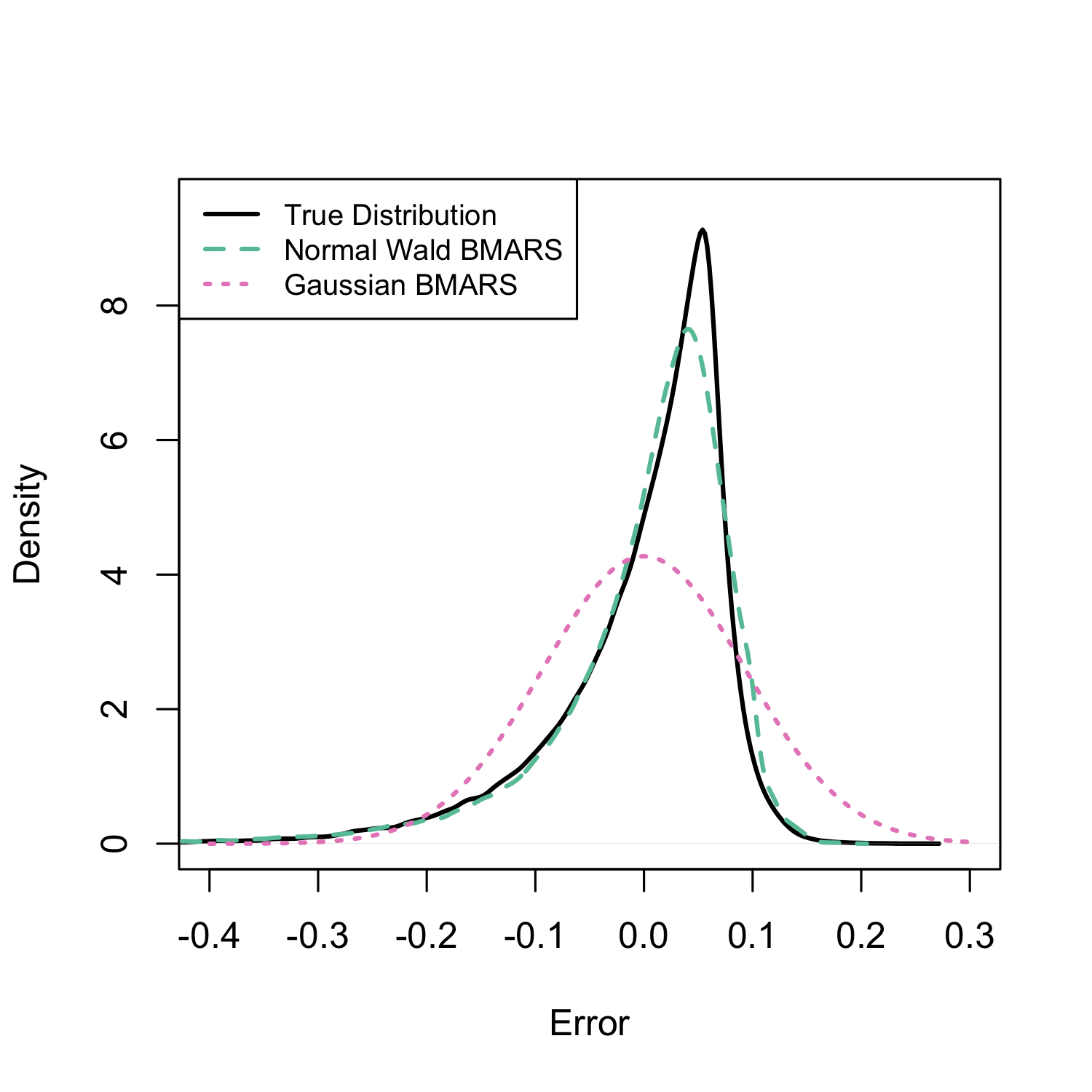}
  \caption{}
  \label{fig:error}
\end{subfigure}%
\begin{subfigure}{.5\textwidth}
  \centering
  \includegraphics[width=.95\linewidth]{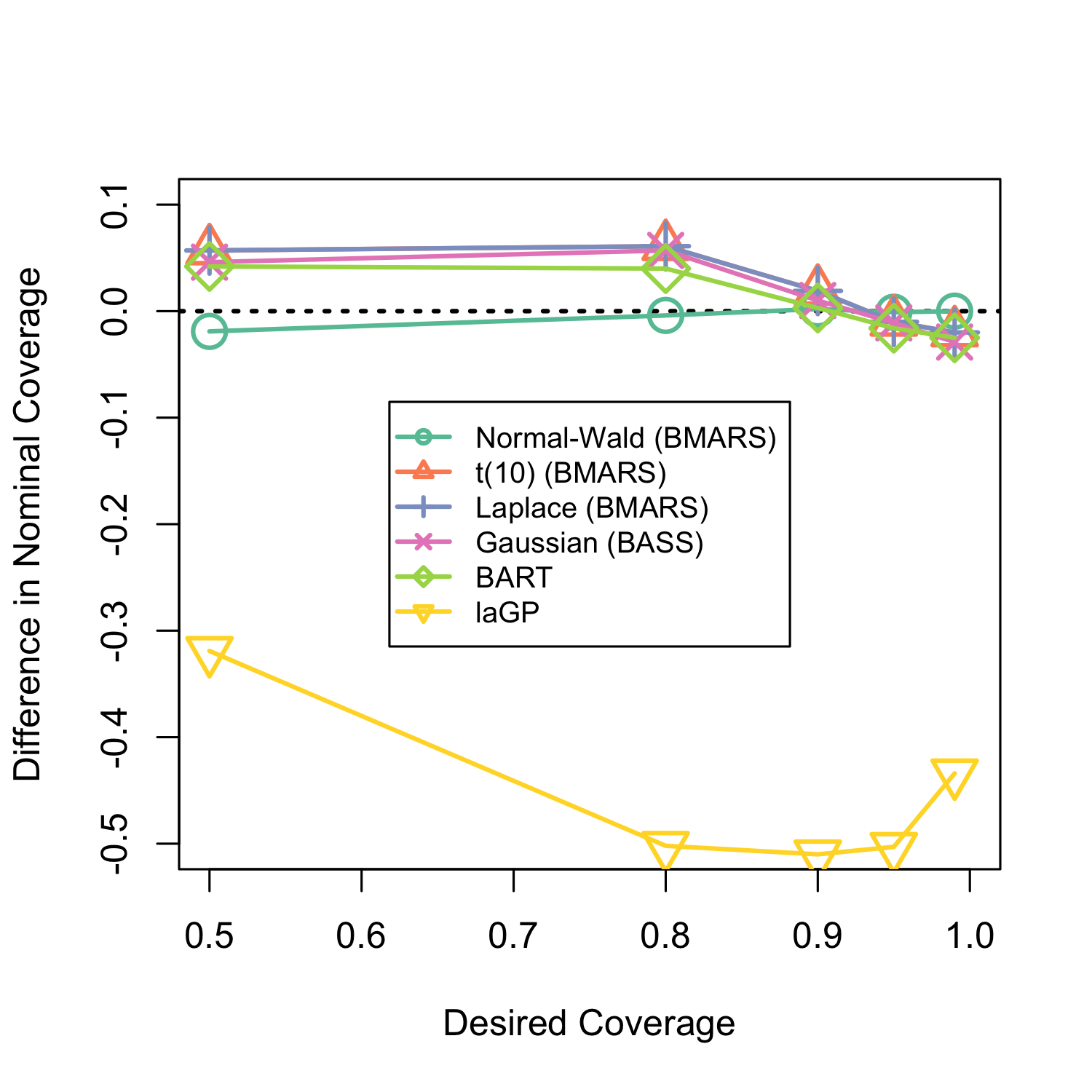}
  \caption{}
  \label{fig:coverage}
\end{subfigure}
\caption{(a) The posterior distribution of the error distribution under Gaussian and $\NW$ likelihoods. GBMARS with a $\NW$ likelihood is able to approximate the shape of the error distribution while simultaneously learning the mean function. (b) Desired coverage versus the difference in nominal coverage for each emulator. Relative to the competitors, GBMARS with a $\NW$ likelihood leads to better calibrated prediction intervals.}
\label{fig:nwbass}
\end{figure}

Next, we train a series of emulators on the training data. We first consider GBMARS with a $\NW$ likelihood. Here, we seek to learn the shape of the error distribution alongside the mean function, leading to better predictive capabilities and uncertainty quantification. We also train a GBMARS with a $t(10)$ likelihood, with the expectation that inference will be more resistant to the non-Gaussian response distribution. The GBMARS emulators are fit using the \texttt{GBASS} package which can be found at \url{https://github.com/knrumsey/GBASS}. We fit a state-of-the-art Gaussian BMARS emulator using the \texttt{BASS} package \citep{francom2020BASS}. We also fit two popular non-MARS emulators including Bayesian additive regression trees (BART; \citep{mcculloch2018bart}) and local approximate Gaussian processes (laGP; \citep{gramacy2016lagp}). All emulators are fit using default settings (for laGP we set \texttt{start=10, end=30}). To compare the performance of the emulators, we generate a test set of $n_T= 1000$ locations using a maximin Latin hypercube design. The metrics computed include (i) root mean square prediction error (RMSPE) between the posterior mean prediction and the error-free output, (ii) average Kolmogorov-Smirnov (KS) distance between the posterior predictive distribution and the true response distribution and (iii) empirical coverage of prediction intervals for $1-\alpha \in \{0.5, 0.8, 0.9, 0.95, 0.99\}$. A full summary of this simulation study is given in \cref{tab:simstudy1}. 

\begin{table}[!htbp] \centering 
  \caption{Simulation study results for piston function with asymmetric Laplace error distribution. GBMARS with a $\NW$ likelihood has the smallest RMSPE and KS distance produces well calibrated prediction intervals.} 
  \label{tab:simstudy1} 
\begin{tabular}{@{\extracolsep{5pt}} lccccccc} 
\\[-1.8ex]\hline 
\hline \\[-1.8ex] 
& & & \multicolumn{5}{c}{Nominal Coverage} \\ \\[-1.8ex]
 & RMSPE  & KS distance & 50\% & 80\% & 90\% & 95\% & 99\% \\ 
\hline \\[-1.8ex] 
$\NW$ & $0.020$ & $0.114$ & $0.481$ & $0.796$ & $0.902$ & $0.949$ & $0.990$ \\ 
$t(10)$ & $0.023$ & $0.143$ & $0.557$ & $0.861$ & $0.919$ & $0.940$ & $0.970$ \\ 
BASS & $0.022$ & $0.157$ & $0.546$ & $0.857$ & $0.910$ & $0.939$ & $0.961$ \\ 
BART & $0.035$ & $0.190$ & $0.542$ & $0.840$ & $0.903$ & $0.934$ & $0.965$ \\ 
laGP & $0.077$ & $0.483$ & $0.181$ & $0.298$ & $0.390$ & $0.447$ & $0.556$ \\ 
\hline \\[-1.8ex] 
\end{tabular} 
\end{table} 

From \cref{tab:simstudy1}, it is clear that GBMARS with a $\NW$ likelihood should be the preferred emulator here. By learning and accounting for the skew and heavy tails of the residual distribution, the inference and uncertainty quantification is better than the Gaussian based alternatives. The $\NW$ case leads to the best predictions (smallest RMSPE) and the best UQ (smallest KS distance) and constructs posterior predictive intervals which are well calibrated across a range of $\alpha$ values (see \cref{fig:coverage}). The calibration of the credible intervals is the direct result of learning the error distribution, as seen in \cref{fig:error}. 

To extend this simulation study, we repeat the above procedure for various error distributions generated using a skewed $t$ distribution \citep{fernandez1998bayesian} with mean $0$ and standard deviation $0.0812$. The inference and UQ for each case and emulator is briefly summarized using RMSPE and KS distance in \cref{tab:simstudy2}. Full simulation results are given in the supplemental materials (SM7). The GBMARS emulator with a $\NW$ likelihood has the smallest KS distance in all $6$ cases, and one of the GBMARS emulators has the lowest RMSPE in $5$ out $6$ cases. Although the $t$ likelihood assumes a symmetric error distribution, it is notably more robust than Gaussian emulators to variations in the true error distribution. 

\begin{table}[!htbp] \centering 
  \caption{Simulation study results for the piston function with a skewed $t$ error distribution (degrees of freedom $\nu$, skew $\kappa$, mean $0$, sd $0.0812$). The bold font indicates the lowest RMSPE or KS distance (KSD) among each of the emulators.} 
  \label{tab:simstudy2} 
\begin{tabular}{@{\extracolsep{5pt}} lrccccc} 
\\[-1.8ex]\hline 
\hline \\[-1.8ex] 
& & $\NW$ & $t(\nu)$  & BASS & BART & laGP \\ 
\hline \\[-1.8ex] 
\multirow{2}{*}{$\nu=4$, $\kappa=-1.25$} & RMSPE & ${\bf 0.023}$ & $0.024$  & $0.037$ & $0.040$ & $0.081$ \\ 
& KSD & ${\bf 0.102}$ & $0.139$ & $0.149$ & $0.188$ & $0.469$ \\ \\[-1.8ex] \\[-1.8ex]
\multirow{2}{*}{$\nu=8$, $\kappa=+1.25$} & RMSPE & $0.022$ & $0.026$  & ${\bf 0.022}$ & $0.033$ & $0.077$ \\ 
& KSD & ${\bf 0.093}$ & $0.117$ & $0.107$ & $0.147$ & $0.472$ \\ \\[-1.8ex] \\[-1.8ex]
\multirow{2}{*}{$\nu=12$, $\kappa=0.00$} & RMSPE & $0.027$ & ${\bf 0.024}$ & $0.027$ & $0.032$ & $0.084$ \\
& KSD & ${\bf 0.084}$ & $0.094$  & $0.095$ & $0.12$ & $0.474$ \\ \\[-1.8ex] \\[-1.8ex]
\multirow{2}{*}{$\nu=20$, $\kappa=-1.00$} & RMSPE & ${\bf 0.021}$ & $0.021$ & $0.029$ & $0.033$ & $0.083$ \\ 
& KSD & ${\bf 0.084}$ & $0.117$  & $0.128$ & $0.160$ & $0.486$ \\ \\[-1.8ex] \\[-1.8ex]
\multirow{2}{*}{$\nu=40$, $\kappa=+0.86$} & RMSPE & $0.024$ & ${\bf 0.023}$ & $0.030$ & $0.030$ & $0.080$ \\ 
& KSD & ${\bf 0.092}$ & $0.104$  & $0.111$ & $0.129$ & $0.469$ \\ \\[-1.8ex] \\[-1.8ex]
\multirow{2}{*}{$\nu=100$, $\kappa=+0.39$} & RMSPE & ${\bf 0.022}$ & $0.026$  & $0.031$ & $0.032$ & $0.078$ \\ 
& KSD & ${\bf 0.086}$ & $0.088$  & $0.096$ & $0.116$ & $0.463$ \\ 
\hline \\[-1.8ex] 
\end{tabular} 
\end{table}

\subsubsection{Complexity Analysis and Scaling Study}
In this subsection, we consider the effect of $n$, the size of the training data, on GBMARS. We will demonstrate that GBMARS shares the convenient scaling properties of BMARS, albeit with substantially larger overhead. Like traditional BMARS, the computational bottleneck of GBMARS comes from obtaining $\bm\Lambda$ (see \cref{eq:accept_gen2}). To compute $\bm\Lambda$, we must invert an $M\times M$ matrix, where $M$ is the number of basis functions in the MARS model at the current MCMC iteration. The number of basis functions required depends on a number of factors including the choice of priors and the behavior of the underlying function, but it generally scales sub-linearly with $n$ and thus BMARS scales favorably with $n$. Unlike BMARS, our generalized Bayesian MARS algorithm requires a Gibbs step for each of the $n$ latent $v_i$ factors, and so the overall complexity of GBMARS is $\mathcal O(n+M^3)$. Additionally, some modeling choices were made in the \texttt{GBASS} package to minimize the memory cost associated with the GBMARS model (which can be substantial in some applications). While these choices do not change the asymptotic complexity of GBMARS, it does lead to a slight increase in the overhead cost of training GBMARS. 

\begin{figure}[t]
    \centering
    \includegraphics[width=0.85\textwidth]{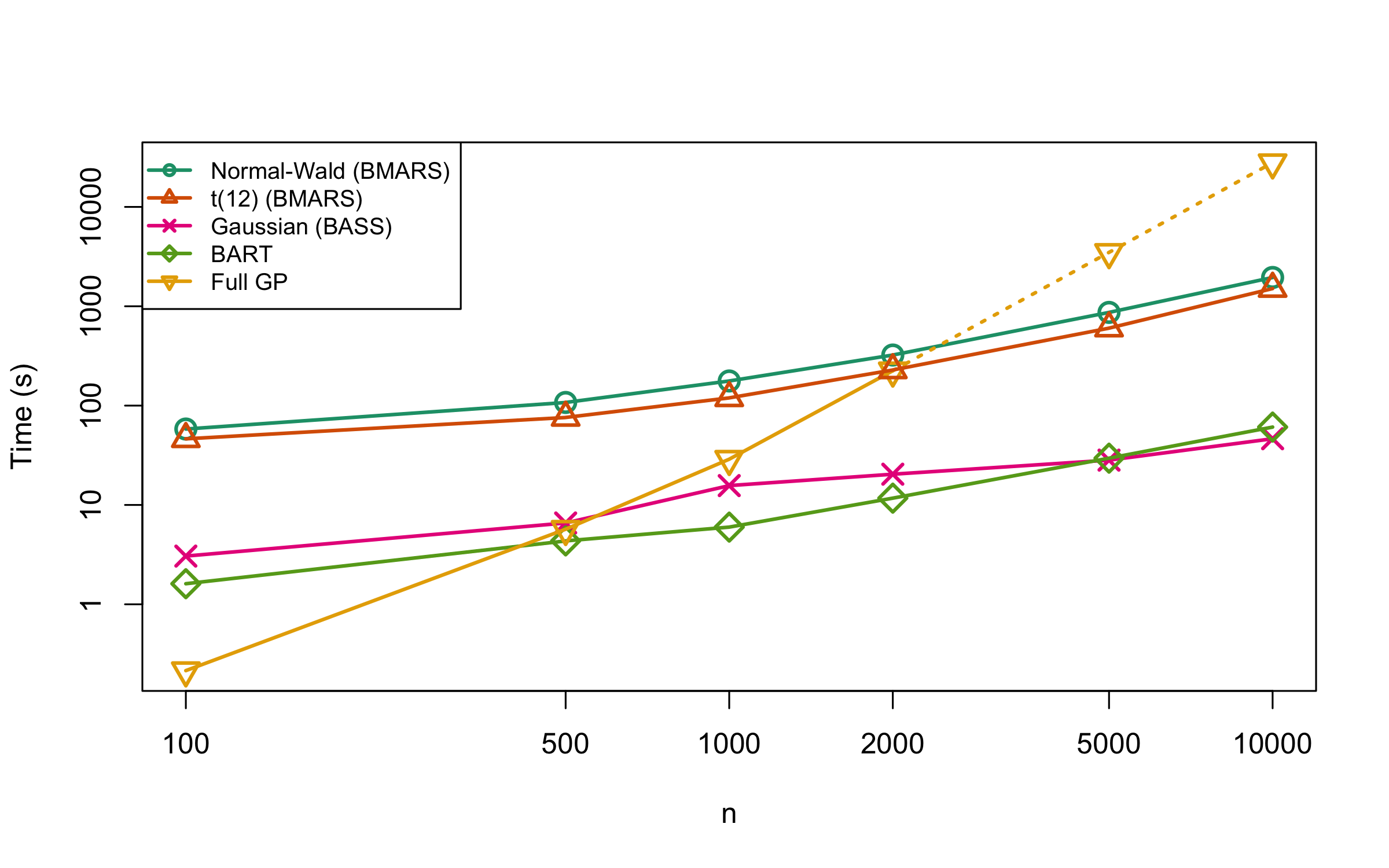}
    \caption{Scaling results for various emulators. Although GBMARS takes longer to train than BASS or BART, the asymptotic behavior is similar and GBMARS is feasible even for very large datasets.}
    \label{fig:scaling}
\end{figure}

To visualize the scaling of GBMARS, we conduct a simple scaling study using the setup described in the previous subsection. In particular, we generate a training set of size $n \in \{100, 500, 1000, 2000, 5000, 10000\}$. The training locations are generated using a random Latin hypercube design and the response is generated as $f(\bm x_i) + \epsilon_i$ where $f()$ is the piston function and $\epsilon_i$ are iid variates from a symmetric $t(12)$ distribution with mean $0$ and  standard deviation $0.0182$. For each training set, we train (i-ii) a GBMARS emulator with a $\NW$ and $t$ likelihood, (iii) a Bayesian MARS emulator with the \texttt{BASS} package, (iv) a BART model with the \texttt{BART} package and (v) a full Gaussian process with the \texttt{mleHomGP} function in the \texttt{hetGP} package. The full GP was fit only for $n \leq 2000$ and the remaining results are extrapolated. The timing results are displayed in \cref{fig:scaling}. Although GBMARS takes longer to train than BMARS or BART, the time costs as a function of $n$ are roughly proportional. This suggests that GBMARS is feasible, even for very large $n$.

\subsection{NetLogo: Quantile Regression Comparison}
In this example, we demonstrate that quantile regression with GBMARS is an excellent tool for the analysis of stochastic computer models. For an additional example, we analyze the $8$ dimensional borehole function with a skewed error distribution in the supplemental materials. We compare the performance of GBMARS to quantile kriging \citep{plumlee2014building} and support vector machines (SVM) \citep{takeuchi2006nonparametric}. Quantile kriging is performed using default settings of the \texttt{quantkriging} package and SVM with the \texttt{qrsvm} package (setting \texttt{cost=1000}). 

\begin{figure}[t]
    \centering
    \includegraphics[width=0.85\textwidth]{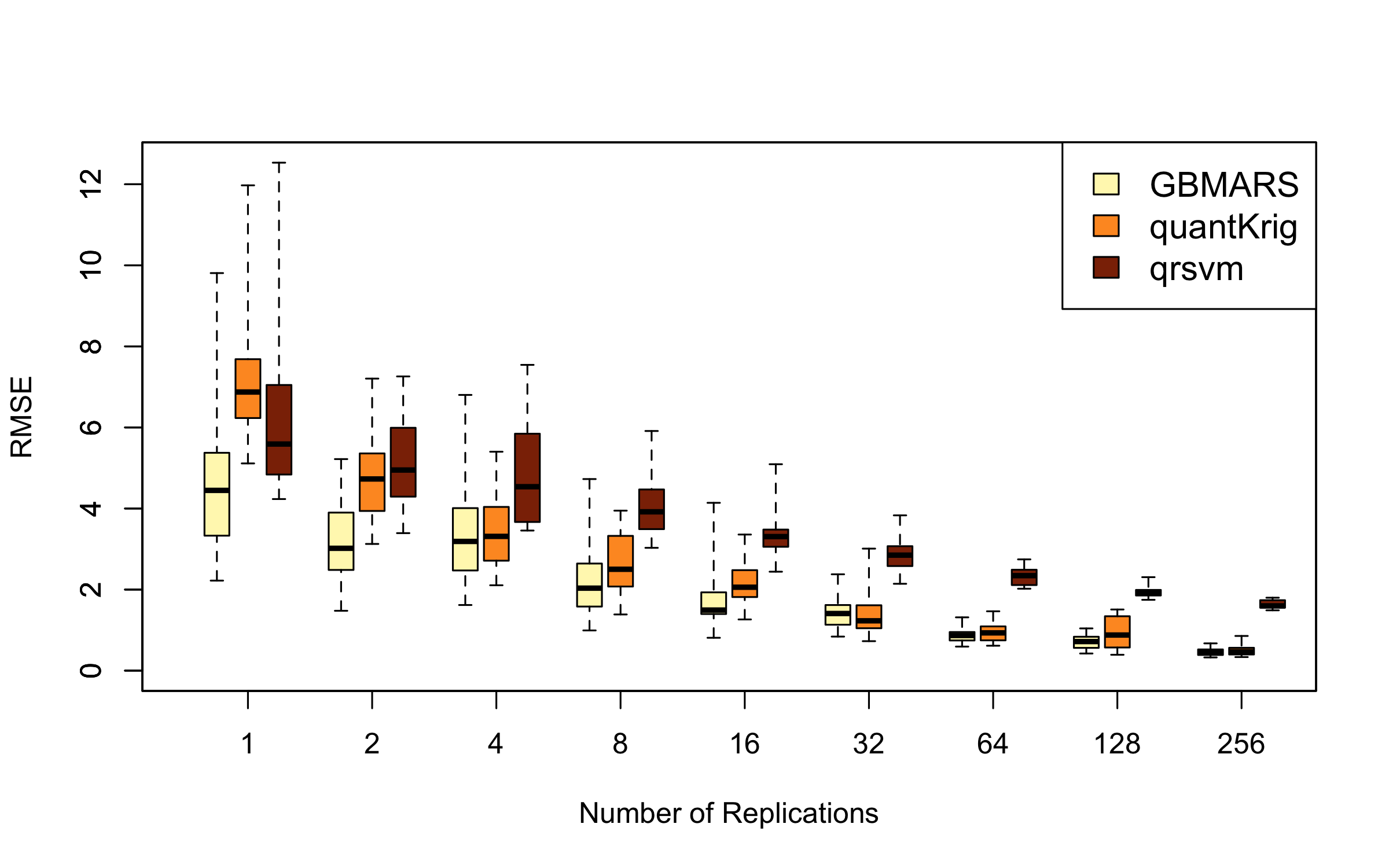}
    \caption{Fish CRC example with different levels of replication. For each case, boxplots give the range of RMSE across $20$ simulated training datasets. Quantile regression with GBMARS is generally more accurate than the alternatives, especially when their are few replications in the training data.}
    \label{fig:fishbox}
\end{figure}

The {\it Capture-Recapture} (CRC) method is an important process in ecology used to approximate the size of a species population. To model this process realistically, one can replace a simple probabilistic model with a sophisticated agent based model, such as the NetLogo model of \cite{wilensky1999netlogo}. For our purposes, we use the dataset described in \cite{baker2022analyzing}. The input is the population of fish in a lake and the model simulates the capturing and tagging of $100$ fish. After a period of time, a second set of $100$ captures are simulated and the number of recaptures is output. For a more detailed explanation of the model, see \cite{baker2020predicting} and the references therein. The dataset contains results for $20$ different population sizes, with $500$ replications for each input. We are interested in the $90^{th}$ percentile of recaptures, conditional on population size. Ground truth is taken to be the conditional sample quantiles of the full dataset. 

Training datasets are constructed by sampling $r$ of the $500$ simulations for each input. Each method is trained on this dataset and predicts the conditional $0.9$ quantile at each of the $20$ locations and the RMSE of these predictions is recorded. This process is repeated $20$ times for each value of $r \in \{2^j\}_{j=0}^8$. The RMSE values are displayed as boxplots in \cref{fig:fishbox}. The percentage of cases for which each method obtained the smallest RMSE is given in \cref{tab:fish}. In this comparison, GBMARS produces the best predictions on average and obtains the smallest RMSE in $68.3\%$ of all cases. The difference in average RMSE is generally largest when there are few replications (small $r$) available. 
\begin{table}[!htbp] \centering 
  \caption{Percentage of the simulations that each method had the smallest RMSE} 
  \label{tab:fish} 
\begin{tabular}{@{\extracolsep{3pt}} lccccccccc|c} 
\\[-1.8ex]\hline 
\hline \\[-1.8ex] 
& \multicolumn{8}{c}{Number of replications} & \\
 & 1 & 2 & 4 & 8 & 16 & 32 & 64 & 128 & 256 & All cases\\ 
\hline \\[-1.8ex] 
GBMARS & $75$ & $80$ & $60$ & $75$ & $85$ & $45$ & $70$ & $60$ & $65$ & $68.3$ \\ 
quantKrig & $10$ & $0$ & $40$ & $25$ & $15$ & $55$ & $30$ & $40$ & $35$ & $27.8$ \\  
qrsvm & $15$ & $20$ & $0$ & $0$ & $0$ & $0$ & $0$ & $0$ & $0$ & $3.9$ \\ 
\hline \\[-1.8ex] 
\end{tabular} 
\end{table}

\subsection{Stochastic SIR Model: A Sensitivity Study}
In this example, we present a novel tool for sensitivity studies using a simple stochastic infectious disease model. 

One benefit of working MARS models is the immediate connection to the Sobol decomposition. Sobol's method is a global approach to sensitivity analysis in which the variance of the response is decomposed and attributed to the individual predictor variables and their interactions \citep{sobol2001global, saltelli2008global}. Like the BMARS framework of \cite{francom2018sensitivity}, GBMARS readily admits closed form Sobol decomposition of the mean function. The unique approach to sensitivity analysis that we propose here, is to fit a sequence of quantile emulators using GBMARS and to examine how the Sobol indices change alongside the quantiles. In this process, we may find evidence of variables which matter relatively more (or less) in the tails of the response distribution, which may be important information for decision making. 

\begin{table}[!htbp] \centering 
  \caption{Description of inputs for the stochastic SIR model.} 
  \label{tab:sir} 
\begin{tabular}{@{\extracolsep{3pt}} lcc} 
\\[-1.8ex]\hline 
\hline \\[-1.8ex] 
Description & Notation & Values \\ 
\hline \\[-1.8ex] 
 Susceptible at $t=0$ (fixed) & $S(0)$ & $5000$ \\
 Infectious at $t=0$ (fixed) & $I(0)$ & $1$ \\
 Recovered at $t=0$ (fixed) & $S(0)$ & $0$ \\ 
 Time of intervention (fixed) & $t_\text{inter}$ & $14$ \\  
\hline \\[-1.8ex] 
Probability of Transmission & $x_1$ & $(0.4, 0.7)$ \\
Average \# of PTIs per person per day & $x_2$ & $(1.0, 4.0)$ \\
Probability of Recovery & $x_3$ & $(0.08, 0.2)$ \\
Efficacy of Intervention & $x_4$ & $(0.5, 0.1)$ \\
Cumulative Infections at day $21$ & y & $\mathbb N_+$ \\
\hline \\[-1.8ex] 
\end{tabular} 
\end{table}

\begin{figure}[ht]
    \centering
    \includegraphics[width=0.95\textwidth]{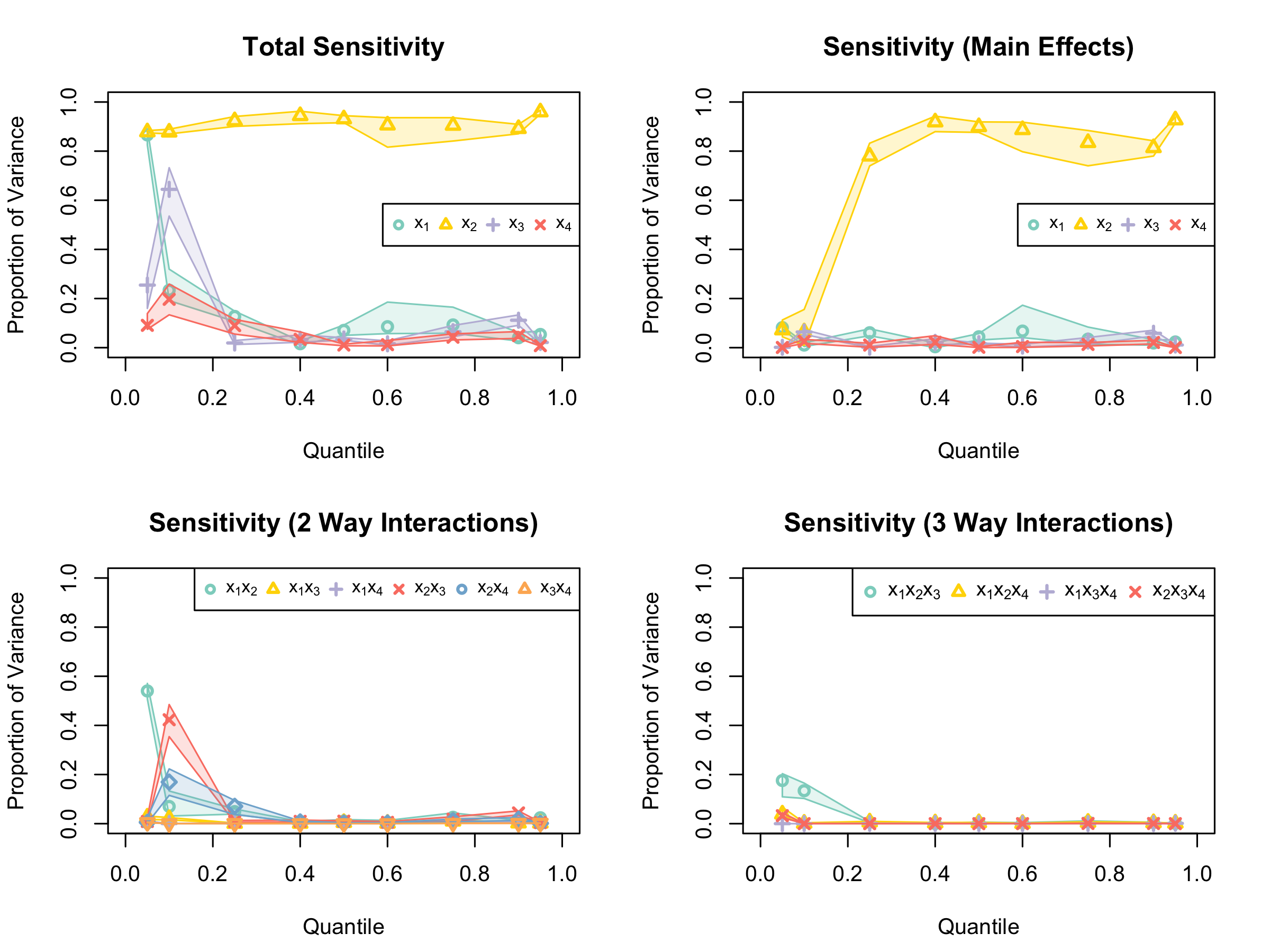}
    \caption{$80\%$ posterior intervals for the Sobol indices of a stochastic SIR model as a function of response quantile. Low quantiles are sensitive to all inputs and their two way interactions, but sensitivity in the large quantiles is dominated by $x_2$.}
    \label{fig:sir}
\end{figure}

To illustrate, we will examine a stochastic susceptible-infectious-recovered (SIR) model which simulates the spread of an infectious disease. Simulations are performed using the \texttt{EpiModel} R package \citep{jenness2018epimodel}. We consider a closed population of $5000$ susceptible people and a single infectious individual on day zero. During each potentially transmissible interaction (PTI), an infectious individual will infect a susceptible individual with probability $x_1$ and all individuals in the population interact at random with an average of $x_2$ PTIs per person per day. Each day, infectious individuals will recover from the disease with probability $x_3$. After $14$ days, an intervention with efficacy $x_4$ is implemented. The response variable is taken to be the cumulative number of infected individuals at the end of a $21$ day period. The relevant simulation inputs, and their ranges, are described in \cref{tab:sir}.

Using a maximin Latin hypercube sample of $2000$ locations, we generate a dataset and we train GBMARS emulators for quantiles $q \in \{0.05, 0.10, 0.25, 0.40, 0.50, 0.60, 0.75, 0.90, 0.95\}$. The Sobol decomposition can be obtained for each GBMARS model (see \cite{francom2018sensitivity} for details). The resulting Sobol indices are plotted, as a function of quantile $q$, in \cref{fig:sir}. In the context of cumulative infections, low and high quantiles can be viewed, respectively, as best case and worst case scenarios. In the best case scenario for the three week development of this disease, which likely corresponds to the non-epidemic or borderline-epidemic case, all of the inputs play an important role in the distribution of the outcome. For example, variance in the $0.1$-quantile of the response distribution is driven by many factors, but especially the interactions between (i) rate of PTIs and recovery speed, (ii) rate of PTIs and intervention efficacy and (ii) rate of PTIs, transmission probability and recovery speed. In the middle and high quantiles (representing most likely and worst case scenarios), sensitivity is driven almost entirely by the rate of PTIs. If this simple stochastic SIR model can be trusted to represent reality, then reducing variability in PTIs will reduce variability in the number of cumulative infections. Given the intuitive monotone relation between PTIs and cumulative infections, this suggests that policies geared towards reducing the rate of PTIs may be most effective strategy for reducing the number of cumulative infections in the worst case scenario.

\section{Conclusion}
\label{sec:conclusion}

In this work, we have introduced a framework for non-linear regression which is of particular interest in the setting of emulation for stochastic computer models. The GBMARS emulator shares several desirable properties with Bayesian MARS, (i) including strong predictive and uncertainty quantification capabilities, (ii) parsimonious and interpretable modeling, (iii) excellent scaling, (iv) connections to popular sensitivity analysis tools and (v) requires little to no tuning. The ability to obtain Bayesian MARS models with respect to the broad class of generalized hyperbolic likelihoods makes it an attractive option, especially for stochastic computer model emulation.

GBMARS with symmetric likelihoods such as the $t$, logistic, variance-gamma and double exponential distributions allow for models which are more robust to the skew and heavy tails commonly associated with stochastic simulators. GBMARS under the flexible $\NW$ likelihood allows for the simultaneous learning of the mean and error structures, leading to improved predictions and better uncertainty quantification (e.g., better calibrated prediction intervals). Finally, quantile regression, a common tool for the analysis of stochastic computer models, with GBMARS is a powerful tool which is competitive with and often superior to popular alternatives (such as quantile kriging). The availability of closed form Sobol decompositions, conditional on a MARS model, makes conducting novel quantile-based sensitivity studies a trivial task. Together, these methods make GBMARS a powerful tool for the analysis of stochastic computer models.

\bibliographystyle{agsm}
\bibliography{references}

\pagebreak
\begin{center}
\textbf{\large Supplemental Materials: Generalized Bayesian MARS: Tools for Emulating Stochastic Computer Models}
\end{center}
\setcounter{equation}{0}
\setcounter{figure}{0}
\setcounter{table}{0}
\setcounter{page}{1}
\setcounter{section}{0}
\makeatletter
\renewcommand{\theequation}{SM\arabic{equation}}
\renewcommand{\thefigure}{SM\arabic{figure}}
\renewcommand{\thesection}{SM\arabic{section}}
\renewcommand{\bibnumfmt}[1]{[SM#1]}
\renewcommand{\citenumfont}[1]{SM#1}

\section{The Conditional Posterior for the Coefficients}
\label{sec:smconditional}

By defining $z_i = y_i - \beta v_i\sqrt w$, the generalized model can be expressed as $z_i = f(\bm x_i) + \epsilon_i, \ \epsilon_i \sim N(0, cwv_i)$, which looks very similar to the BMARS model of Equation (1). In matrix form, we can write the GBMARS model as
\begin{equation}
\label{eq:smgbmars}
    \begin{aligned}
    \bm z|\cdot &\sim N\left(\bm B \bm a, cw\bm V\right) \\
    \bm a|w,\tau,\bm\Sigma &\sim N\left(\bm 0, w\tau \bm\Sigma\right), \\
    \tau &\sim \text{InvGamma}(a_\tau, b_\tau), \\
    M &\sim \pi_M, \quad
    \bm\theta_B \sim \pi_B, \quad
    w \sim \pi_w, \quad
    v_i \stackrel{\text{iid}}{\sim} \pi_v, \quad
    \beta \sim \pi_\beta
    \end{aligned}
\end{equation}
where $\bm V$ is a diagonal matrix whose $(ii)^{th}$ component is $v_i$. We will sample from this model using an algorithm which is nearly identical to the one described in Section 2, and any departures will be discussed. This latent variable structure is useful because the coefficients can still be integrated out of the posterior. To see this, note that 
\begin{equation}
\label{eq:smgbmars_prod}
\begin{aligned}
N(\bm z | \bm B \bm a, cw\bm V)N(\bm a &| \bm 0, w\tau\bm\Sigma) = \\
&\left(\frac{w^N \left\{\prod_{i=1}^Nv_i\right\} \tau^{M+1}}{c^{M+1}}\frac{|\bm\Sigma|}{|\bm\Lambda|}\right)^{-1/2} \\
& \times  \ \exp\left\{-\frac{1}{2wc}\left(\sum_{i=1}^N\frac{z_i^2}{v_i} - \bm z' \bm V^{-1}\bm B'\bm\Lambda\bm B'\bm V^{-1}\bm z\right) \right\} \\
&\times \ N\left(\bm a \ \bigg| \ \bm\Lambda\bm B'\bm V^{-1}\bm z, \ wc\bm\Lambda\right),
\end{aligned}
\end{equation}
where $\bm\Lambda = \left(\bm B'\bm V^{-1}\bm B + \frac{c}{\tau}\bm\Sigma^{-1}\right)^{-1}$. 

\section{The Generalized Beta Prime Prior}
\label{sec:smGBP}

In some cases, it is useful to examine the quantity $\kappa_i = \frac{1}{1+wv_i}$ which we will refer to as an {\it interpolative coefficient}. Values of $\kappa_i \approx 1$ indicate that the $i^{th}$ data point has been nearly interpolated, while values of $\kappa_i \approx 0$ indicate a large amount of variance. These quantities have been thoroughly studied in the field of Bayesian regularization \citep{carvalho2009}, where they are referred to as shrinkage coefficients. In Bayesian regularization, these relate directly to the coefficients of a regression model. We propose to leverage these ideas in a different way, relating them instead to the error of each measurement. 

In the case of regression with possible outliers, it is reasonable to desire a prior distribution for $\kappa_i$ with significant mass near both $0$ and $1$. This principle, which can be summarized as ``interpolate or ignore'', implies that the error should be as small as possible on one hand ($\kappa_i \approx 1$). On the other hand, if a data point is identified as a possible outlier ($\kappa_i \approx 0)$, we want to associate with it a large variance, to avoid corrupting the model fit. In the realm of Bayesian regularization, priors are often considered directly in terms of these coefficients. Two popular choices of prior (i) the Horseshoe and (ii) the Strawderman-Berger prior can be realized as placing a Beta prior directly on the coefficients for $\kappa_i$. The Horseshoe prior corresponds to $\kappa_i \sim \text{Beta}(1/2, 1/2)$ and the Strawderman-Berger prior corresponds to $\kappa_i \sim \text{Beta}(1, 1/2)$. After transformation, these priors can be unified as specifying a generalized Beta prime \citep{mcdonald1995generalization} prior for each $v_i$. The GBP prior has density
$$\text{GBP}(x|p,a,b) \propto x^{ap-1}(1+x^p)^{-a-b}, \ x > 0, \ p,a,b > 0.$$
In this setting, the Horseshoe prior corresponds to $p=1, a=1/2, b=1/2$ and the Strawderman-Berger prior corresponds to $p=1/2, a=1, b=1/2$. In practice, this prior does not lead to a tractable full conditional, and we use a Metropolis-Hastings procedure to sample from the full conditional posterior. 

Another advantage of the GBP prior is that the special case GBP($1, 1/2, 1/2)$ is equivalent to specifying a Half-Cauchy prior for the scale, and is often a better default choice for the global variance factor $w$ than Jeffreys' prior \citep{gelman2006prior}. Since these priors are not scale invariant, we propose setting the fixed variance factor $c$ equal to the variance of the data. 

With this GBMARS framework with GIG or GBP priors for local variance factors in place, we are ready to demonstrate the variety of modeling situations in which GBMARS can be applied.

\section{Setting Hyperparameters for Tail Heaviness}
\label{sec:smhyper}

Recall that the prior for $\gamma$ is of the form 
$$\gamma \sim N\left(m_\gamma, s_\gamma^2\right).$$
The relationship between $\xi$ and $\gamma$ can be written as
\begin{equation}
    \label{eq:smxi}
    |\gamma| = \frac{1}{\xi^2} - 1.
\end{equation}
The LHS of the above equation must have a folded normal distribution, with distribution function
$$F_{|\gamma|}(x) = \frac{1}{2}\left(\text{erf}\left(\frac{x+\mu}{\sqrt 2\sigma} \right) + \text{erf}\left(\frac{x-\mu}{\sqrt 2\sigma}\right) \right)$$
where $\text{erf}()$ denotes the error function.

If $\xi \sim \text{Beta}(a, b)$, then the distribution function for the RHS is
$$F_{\xi^{-2}-1}(x) = 1 - I_{\sqrt{1/(x+1)}}(a, b),$$
where $I_x(a, b)$ is the regularized incomplete Beta function. Strictly speaking, there is no way to specify a Beta distribution for $\xi$ so that \cref{eq:smxi} holds in distribution. Given $a$ and $b$, however, we can choose corresponding values of $m_\gamma$ and $s_\gamma$ that minimizes the distance between these two distribution functions in some sense. $L_2$ and Kolmogorov Smirnov distance are two reasonable choices for doing this, but an easy alternative is to simply find the values of $m_\gamma$, $s_\gamma$ for which the distribution functions are equal to each other for $2$ different values. This leads to a fast-to-evaluate objective function which can easily be optimized with a variety of optimization algorithms, such as L-BFGS-B \citep{zhu1997algorithm}, which can be implemented using the \texttt{optim} function in base R. 

To recover the values given in the main manuscript, we find the values of $m_\gamma$ and $s_\gamma$ so that the two distribution functions have the same input for values $0.1$ and $0.9$. 

\section{Robust Regression in the Presence of Outliers}
\label{sec:smoutliers}

Consider $n$ data points which are observed according to the process $y_i = f(\bm x_i) + \epsilon_i + \delta_i$. The $\epsilon_i$ are iid error components with common variance $\sigma^2$, and $\delta_i$ is a corruption mechanism such that
$$
\delta_i \sim \begin{cases}
N(0, \sigma_\delta^2), & i \in \mathcal I \\
0, & i \notin \mathcal I, 
\end{cases}
$$
for some $\mathcal I \subset \{1,2,\ldots n\}$. The primary objectives are (i) to learn the underlying function $f(\cdot)$, (ii) to infer the variance of the uncorrupted data $\sigma^2$, and (iii) identify the outliers. The standard approach to BMARS can be hyper-sensitive to the corrupted data when $\sigma_\delta$ and/or $|\mathcal I|$ is large, leading to poor inference for one or both of these objectives.  This is a critique of many machine learning techniques more generally, that outliers can have devastating effects. By adjusting the likelihood function, the GBMARS framework of \cref{sec:GBMARS} can lead to improved prediction (better inference for $f(\cdot)$) and better quantification of uncertainty (better inference for $\sigma^2$). There are many interesting choices of likelihood function for this problem, but we will focus on the $t$-distribution and the Horseshoe likelihood. 

\begin{figure}[htbp]
\centering
\includegraphics[width=0.75\textwidth]{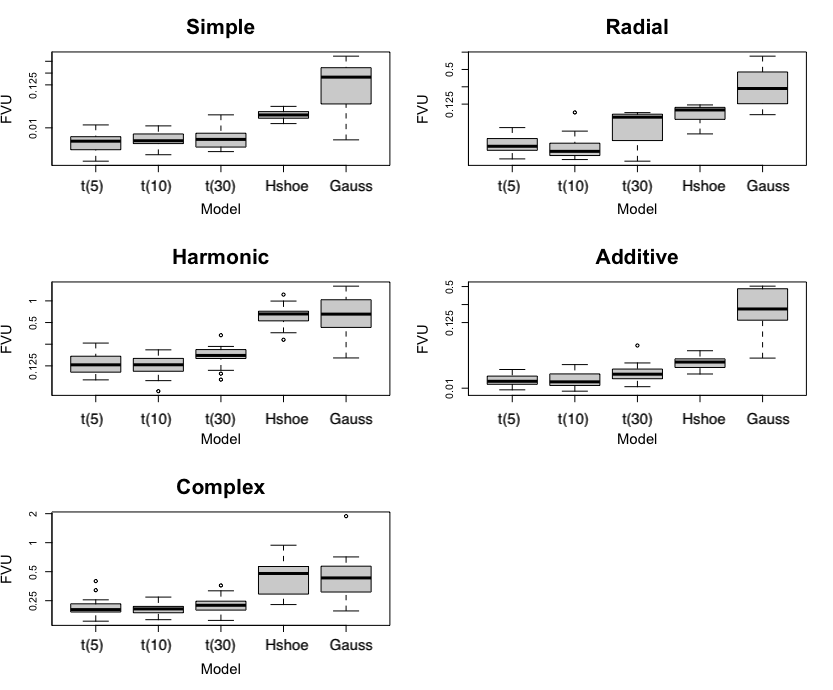}
\caption{A comparison of GBMARS with 5 different likelihoods when outliers are present.  Each plot shows 5 boxplots that give the out-of-sample FVU (log scale) for the GBMARS models under the 5 likelihoods.  The boxplots are based on 30 repetitions of the data generation and model fitting. }
\label{fig:smboxplots_fvu}
\end{figure}

To assess the performance of these likelihoods, we will adapt the simulation study of \cite{denison1998} and \citep{nott2005}. We begin by considering five different test functions, which are referred to as (1) simple interaction, (2) radial, (3) harmonic, (4) additive and (5) complex interaction respectively, 
\begin{align*}
f_1(\bm x) &= 10.391((x_1 - 0.4)(x_2-0.6)+0.36) \\[1.1ex]
f_2(\bm x) &= 24.234(r^2(0.75-r^2)), \ r^2 = (x_1-0.5)^2 + (x_2-0.5)^2 \\[1.1ex]
f_3(\bm x) &= 42.659(0.1 + \hat x_1(0.05+\hat x_1^4 - 10\hat x_1^2\hat x_2^2 + 5\hat x_2^4)), \ \hat x_i=x_i-0.5 \\[1.1ex]
f_4(\bm x) &= 1.3356(1.5(1-x_1)+\exp(2x_1-1)\sin(3\pi(x_1-0.6)^2)+\\[1.1ex]
&\hspace{1in}\exp(3(x_2-0.5))\sin(4\pi(x_2-0.9)^2)) \\[1.1ex]
f_5(\bm x) &= 1.9135 + \exp(x_1)\sin(13(x_1-0.6)^2)\exp(-x_2)\sin(7x_2). \\[1.1ex]
\end{align*}
Although each of these functions is influenced by only two variables, we add three additional inert predictors into the mix. We begin by generating predictor variables $\bm x_i$ ($i=1,2,\ldots 225)$ using a space-filling design over the space $[0,1]^5$. For each test function $(j=1,\ldots 5)$, we simulate data as
\begin{align*}
y_i &= f_j(\bm x_i) + \epsilon_i + \delta_i \\
\epsilon_i &\stackrel{\text{iid}}{\sim} N(0, 0.25^2) \\
\delta_i &\sim \begin{cases}
N(0, 4.75^2), & i=1,2,\ldots 5 \\
0, & i = 6, 7, \ldots 225
\end{cases}
\end{align*}
The first $5$ observations are then corrupted by adding Gaussian noise to these observations with a standard deviation of $4.75$. 

\begin{figure}[htbp]
\centering
\includegraphics[width=0.75\textwidth]{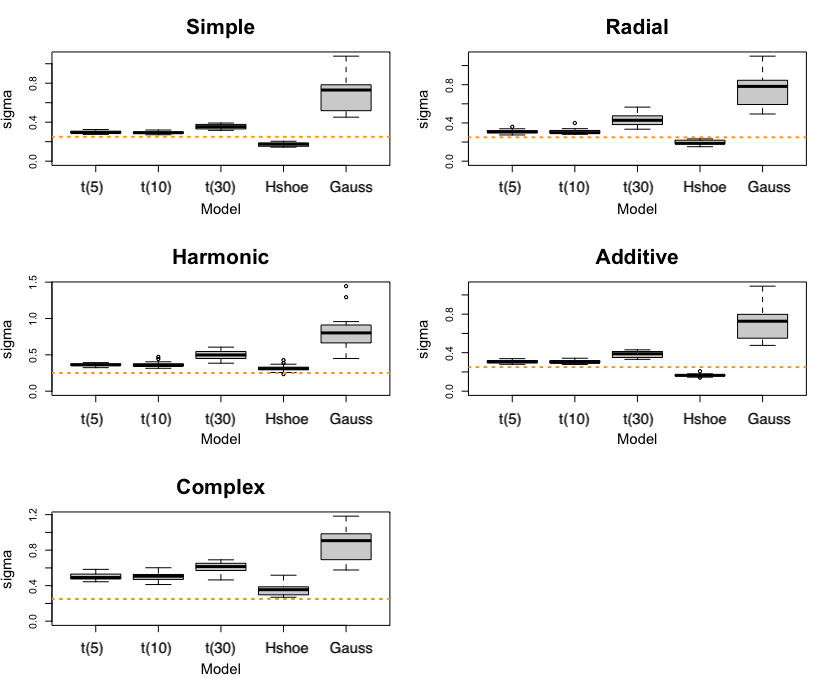}
\caption{A comparison of GBMARS with 5 different likelihoods when outliers are present.  Each plot shows 5 boxplots that give the global variance factor for the GBMARS models under the 5 likelihoods.  The boxplots are based on 30 repetitions of the data generation and model fitting.  The true (outlier free) value is shown with a horizontal line.}
\label{fig:smboxplots_s2}
\end{figure}

Using default settings, we fit $5$ GBMARS models using a (i-iii) $t$ likelihood with $5$, $10$ and $30$ degrees of freedom, (iv) a Horseshoe likelihood and (v) a normal likelihood (using the default settings described in \cite{francom2020}). Following \cite{nott2005}, we measure the predictive capabilities of each model using the fraction of variance unexplained (FVU), which is estimated out-of-sample using $10000$ Monte Carlo replicates.  As seen in \cref{fig:smboxplots_fvu}, the $t$ likelihood facilitates the best inference for the underlying function, yielding the smallest FVU values. The Horseshoe likelihood outperforms the normal likelihood for some of the test functions, but is generally inferior to the $t$ models. \cref{fig:smboxplots_s2} shows the posterior estimate of the standard deviation for each of the thirty runs, where the horizontal dashed line represents the target $\sigma = 0.25$. It is apparent that the corrupted data is heavily influencing the fit under the normal likelihood, leading to poor predictive performance and overestimation of the variance. The $t$ and Horseshoe priors are far more robust to outliers.  

Note that the GBMARS framework is not significantly more expensive than the BMARS framework, meaning that the algorithm scales roughly linearly in sample size and that for moderate sample sizes we can obtain tens of thousands of MCMC iterations in a matter of minutes.

\section{Quantile Regression in High Dimensions}
\label{sec:smborehole}

\begin{figure}[htbp]
\centering
\includegraphics[width=0.75\textwidth]{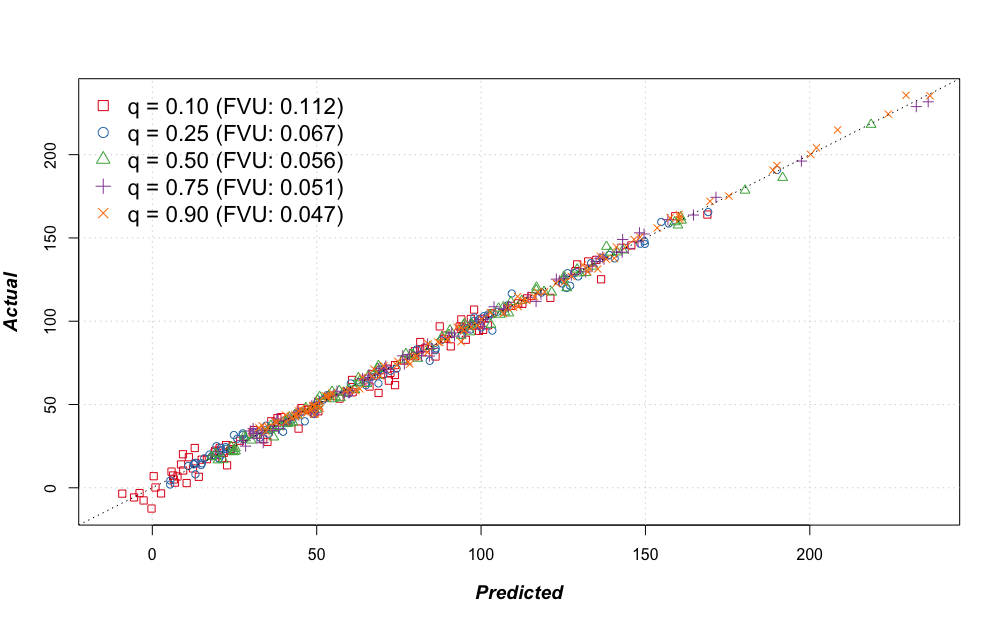}
\caption{GBMARS quantile regression posterior mean predictions versus actual quantile values under the stochastic borehole example.}
\label{fig:smqline}
\end{figure}

In this section, we explore quantile regression using GBMARS in a higher dimensional problem. The training data consists of $5000$ stochastic realizations of the Borehole function
\begin{equation}
    \label{eq:smborehole}
    f(\bm x) = \frac{2\pi x_3(x_4-x_6)}{\log(x_2/x_1)\left(1 + \frac{2x_7x_3}{\log(x_2/x_1)x_1^2x_8} + \frac{x_3}{x_5}\right)}.
\end{equation}
We use a Latin hypercube sample to generate a training set using the standard input ranges for each of the $8$ variables (see \cite{surjanovic2013virtual}). The response variable is generated as
$y_i = f(\bm x_i) + \epsilon_i, \ i=1,2\ldots 5000$
where each $\epsilon_i$ is a three parameter Weibull random variable with mean zero, shape $20$ and scale $250$. This is a left skewed error model leading to a signal-noise ratio of roughly $10$. This form is convenient, because the closed form quantile function makes it easy to assess the accuracy of each quantile regression. We perform quantile regression for five quantiles, $q=0.10, 0.25, 0.50, 0.75$ and $0.90$. A second Latin hypercube design consisting of $100$ points was obtained for testing, and the model predictions (posterior mean) were compared to ground truth for each of the quantiles. \cref{fig:smqline} shows the predicted vs actual values, and the FVU for each quantile is included in the legend. All of the FVU values are quite low, with the largest discrepancy occurring for the $0.10$ quantile. Due to the left skew of the error model, this is expected since the target is in the longer tail of the distribution. 

\section{Applications}
\label{sec:smapplications}

In this section, we will compare the performance of BMARS with (i) Gaussian (ii) t($5$) and (iii) $\NIG$ likelihoods on seven data sets of varying sizes. The data sets we explore are briefly described as follows. 

\begin{figure}[htbp]
\centering
\includegraphics[width=0.75\textwidth]{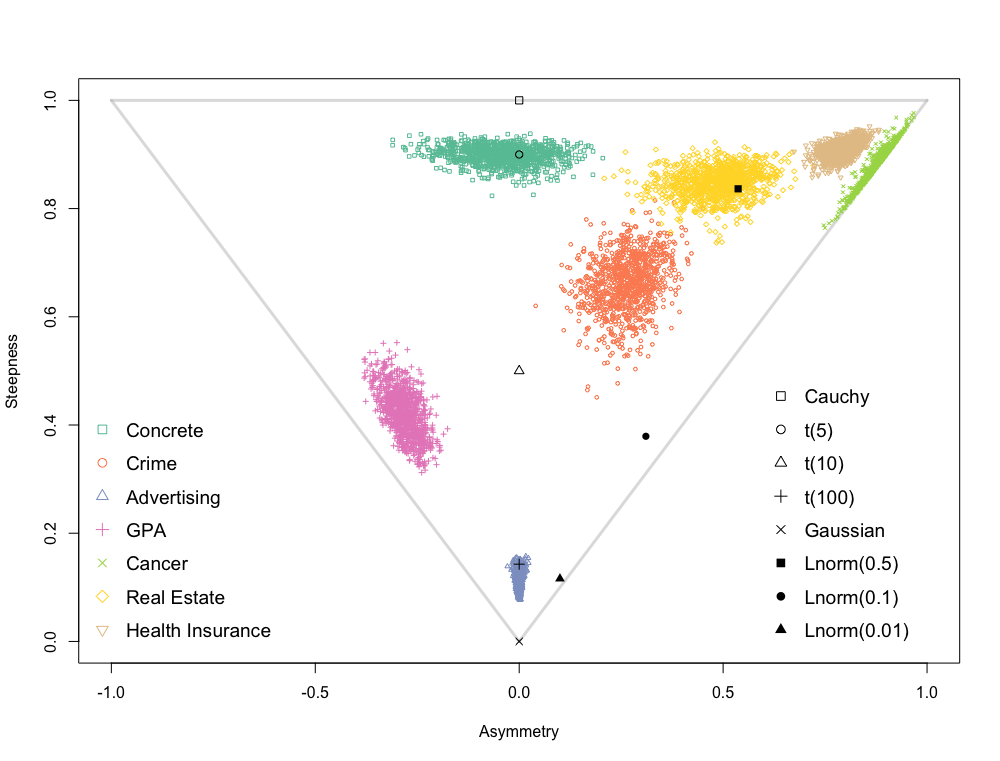}
\caption{$\NIG$ triangle plot showing asymmetry ($\chi$) and steepness ($\xi$) on the $x$ and $y$ axes, respectively.  A few reference distributions are shown, along with posterior samples for GBMARS $\NIG$ regression models trained using seven example data sets.}
\label{fig:smtriangle}
\end{figure}

\begin{enumerate}
    \item {\it Concrete:} We attempt to model the compressive strength of concrete, as a function of $8$ covariates describing the age and composition of the concrete, with $n=1030$ \citep{yeh1998modeling}.
    \item {\it Crime:} This data set contains demographic information about the $440$ largest U.S. counties. We are interested in modeling the number of annual serious crimes per capita using $11$ covariates \citep{kutner2005applied}. 
    \item {\it Advertising:} We aim to predict the market share of a product, using $n=36$ observations and $4$ covariates describing the advertising strategy \citep{kutner2005applied}. 
    \item {\it GPA:} We seek to understand the GPA of $705$ applicants to a state university (between $1996$ and $2000$) using high school class rank, ACT score and Academic year \citep{kutner2005applied}. 
    \item {\it Cancer:} This data set contains clinical measurements for $97$ men with advanced prostate cancer. The goal is to understand the relationship between prostate-specific antigen level and $7$ additional prognostic measurements \citep{kutner2005applied}. 
    \item {\it Real Estate:} We seek to predict the sales price of a residential home (midwestern city, year $2002$) based on $10$ quantitative (or binary) characteristics of the home and surrounding property ($n=522$) \citep{kutner2005applied}. 
    \item {\it Health Insurance:} This data set contains information on $788$ health insurance subscribers who made heart disease claims. We seek to predict the total cost of services provided based on $7$ covariates. Due to the right skew of the response, the square root of the total cost is modeled instead \citep{kutner2005applied}. 
\end{enumerate}

One nice feature about GBMARS with an $\NIG$ likelihood is the ability to visualize the data set using the $\NIG$ triangle (see Section 6). In \cref{fig:smtriangle}, the posterior samples of steepness and asymmetry $(\chi, \xi)$ are plotted for each data set. For reference, we also show the location in the $\NIG$ triangle of several $t(\nu)$ and log-normal($\sigma$) distributions. \Cref{fig:smtriangle} is helpful in its own right, as it helps to inform the appropriate likelihood for each data set. For example, the $\NIG$ triangle plot indicates that a $t$ distribution with $5$ degrees of freedom may be appropriate for the concrete data set, and the standard Gaussian likelihood is likely to be appropriate for the advertising data set. For the remaining applications, a more flexible approach, such as $\NIG$ or quantile regression may be needed. An alternative solution is to attempt to transform the response value so that it becomes approximately Gaussian, or at least symmetric. For instance, the Real Estate data set may become approximately normal after a log transformation. This approach has drawbacks in certain settings, such as the limited interpretability of the Sobol decomposition. 

For each data set, we would like to compare the behavior and performance of GBMARS under three different likelihoods: (i) Gaussian, (ii) $t$ with $5$ degrees of freedom and (iii) $\NIG$. To test the out-of-sample capabilities of each model, we perform $10$ fold cross validation (CV) for each data set and we calculate (i) the FVU, (ii) empirical coverage of the $90\%$ predictive credible interval and (iii) average interval length. Using the model trained on the full data, we also compute a marginalized form of Bayesian information criteria (BIC), a (non-Bayesian) goodness of fit metric which penalizes $\NIG$ for having extra parameters. We note that all three likelihoods can be viewed as special cases of the generalized hyperbolic distribution, and can therefore be interpreted nested models. Since the goal is to estimate the regression function as well as possible, we do not want to penalize a model for the number of basis functions it selects. Therefore we marginalize over all parameters except for those which define the residual distribution: $w$ for the first two models and $w,\gamma,\beta$ for the $\NIG$ model. We begin by obtaining posterior estimates for $\bm{\hat y}$, $w$ (and $\beta,\gamma$ where applicable) for each model, and we compute
\begin{equation}
BIC = k\log(n) - 2\sum_{i=1}^n\log f(y_i - \hat y_i| \text{estimated parameters}),
\end{equation}
where $n$ is the sample size, $k$ is the number of estimated parameters ($k=1$ for Gaussian/$t(5)$ and $k=3$ for $\NIG$) and $f(\cdot)$ represents the appropriate residual density for each model. This metric assesses the overall fit of the model, both in terms of shape and precision. The results are reported in Table 2. 

\begin{table} 
\caption{Comparison of three likelihoods for each of seven data sets.  The out-of-sample FVU, empirical coverage of 90\% intervals, and interval widths are compared as well as the in-sample BIC. }
  \centering 
\begin{tabular}{@{\extracolsep{3pt}} llccccccc} 
\\[-1.8ex]\hline 
\hline \\[-1.8ex] 
& Model & Concrete & Crime & Advert. & GPA & Cancer & R.E. & Insurance \\
\hline \\[-1.8ex] 
 & Gauss  & $0.420$ & $0.748$ & $0.414$ & $0.999$ & $0.993$ & $0.237$ & $0.374$ \\ 
FVU & t(5)  & $0.265$ & $0.684$ & $0.424$ & $0.935$ & $0.99$ & $0.265$ & $0.425$ \\ 
 & $\NIG$  & $0.254$ & $0.69$ & $0.415$ & $0.972$ & $0.971$ & $0.204$ & $0.471$ \\ 
 \hline \\[-1.8ex] 
 & Gauss  & $0.741$ & $0.857$ & $0.917$ & $0.796$ & $0.753$ & $0.847$ & $0.907$ \\ 
Cov. & t(5)  & $0.700$ & $0.998$ & $1.000$ & $0.722$ & $0.753$ & $0.805$ & $0.86$ \\ 
 & $\NIG$  & $0.694$ & $0.866$ & $0.889$ & $0.722$ & $0.835$ & $0.83$ & $0.86$ \\ 
 \hline \\[-1.8ex] 
 & Gauss  & $15.87$ & $0.057$ & $0.575$ & $1.82$ & $37.243$ & $0.172$ & $74.101$ \\ 
Wid. & t(5)  & $13.447$ & $0.162$ & $0.686$ & $1.845$ & $34.425$ & $0.153$ & $48.881$ \\ 
 & $\NIG$  & $13.817$ & $0.057$ & $0.568$ & $1.712$ & $50.715$ & $0.147$ & $49.831$ \\ 
 \hline \\[-1.8ex] 
 & Gauss  & $6071$ & $-2338$ & \bm{$-26.64$} & $1171$ & \bm{$637.7$} & $12802$ & $7059$ \\ 
BIC & t(5)  & \bm{$5577$} & $-1884$ & $-25.70$ & $1161$ & $784.6$ & $12683$ & $6693$ \\ 
 & $\NIG$  & $5943$ & \bm{$-2339$} & $-19.87$ & \bm{$1127$} & $719.4$ & \bm{$12640$} & \bm{$6454$} \\ 
 \hline \\[-1.8ex]
\end{tabular} 
\end{table} 

We begin our discussion with the advertising data set, which has nearly Gaussian residuals according to \cref{fig:smtriangle}. Unsurprisingly, the Gaussian GBMARS model is the best performer according to BIC and FVU, although we note that $\NIG$ obtains a similar level of coverage, with similar (but slightly more precise) interval width. Similarly, \cref{fig:smtriangle} indicates that the $t(5)$ model should be the winner for the concrete data, and again, we find this to be the case. It is worth noting that the $\NIG$ model, which is capable of modeling $t$ distributions, outperforms the Gaussian model here. In terms of BIC, the $\NIG$ model is the best for four of the remaining five data sets. The Gaussian model has the lowest BIC for the cancer data, but the other metrics arguably favor $\NIG$. Although the credible interval width is significantly larger for $\NIG$, the coverage is much better for this model, indicating that the Gaussian and $t(5)$ models are struggling to adequately capture the large right skew of these residuals. 

\section{Piston Function with Skewed t Error}
\label{sec:smskewt}

Here, we provide tables giving the full output of the piston simulation study conducted in Section 4. Tables constructed using the \texttt{stargazer} package in R \citep{hlavac2018stargazer}.

\begin{table}[!htbp] \centering 
  \caption{$\nu = 4, \kappa=-1.25$} 
  \label{tab:smpiston1} 
\begin{tabular}{@{\extracolsep{5pt}} cccccccc} 
\\[-1.8ex]\hline 
\hline \\[-1.8ex] 
 & RMSE\textasteriskcentered  & KS-dist & Cov 0.5 & Cov 0.8 & Cov 0.9 & Cov 0.95 & Cov 0.99 \\ 
\hline \\[-1.8ex] 
GBASS & $0.021$ & $0.092$ & $0.471$ & $0.784$ & $0.902$ & $0.955$ & $0.994$ \\ 
TBASS & $0.022$ & $0.116$ & $0.533$ & $0.856$ & $0.936$ & $0.970$ & $0.991$ \\ 
BASS & $0.026$ & $0.115$ & $0.496$ & $0.827$ & $0.909$ & $0.954$ & $0.978$ \\ 
BART & $0.032$ & $0.147$ & $0.507$ & $0.824$ & $0.914$ & $0.945$ & $0.978$ \\ 
laGP & $0.079$ & $0.479$ & $0.152$ & $0.284$ & $0.365$ & $0.434$ & $0.563$ \\ 
\hline \\[-1.8ex] 
\end{tabular} 
\end{table} 

\begin{table}[!htbp] \centering 
  \caption{$\nu = 8, \kappa=1.25$} 
  \label{tab:smpiston2} 
\begin{tabular}{@{\extracolsep{5pt}} cccccccc} 
\\[-1.8ex]\hline 
\hline \\[-1.8ex] 
 & RMSE\textasteriskcentered  & KS-dist & Cov 0.5 & Cov 0.8 & Cov 0.9 & Cov 0.95 & Cov 0.99 \\ 
\hline \\[-1.8ex] 
GBASS & $0.022$ & $0.074$ & $0.503$ & $0.824$ & $0.919$ & $0.971$ & $0.996$ \\ 
TBASS & $0.022$ & $0.087$ & $0.574$ & $0.857$ & $0.928$ & $0.967$ & $0.991$ \\ 
BASS & $0.027$ & $0.096$ & $0.538$ & $0.837$ & $0.907$ & $0.945$ & $0.980$ \\ 
BART & $0.032$ & $0.114$ & $0.515$ & $0.824$ & $0.918$ & $0.948$ & $0.985$ \\ 
laGP & $0.080$ & $0.484$ & $0.159$ & $0.305$ & $0.395$ & $0.463$ & $0.575$ \\ 
\hline \\[-1.8ex] 
\end{tabular} 
\end{table} 

\begin{table}[!htbp] \centering 
  \caption{$\nu = 12, \kappa=0.00$} 
  \label{tab:smpiston3} 
\begin{tabular}{@{\extracolsep{5pt}} cccccccc} 
\\[-1.8ex]\hline 
\hline \\[-1.8ex] 
 & RMSE\textasteriskcentered  & KS-dist & Cov 0.5 & Cov 0.8 & Cov 0.9 & Cov 0.95 & Cov 0.99 \\ 
\hline \\[-1.8ex] 
GBASS & $0.023$ & $0.099$ & $0.458$ & $0.800$ & $0.910$ & $0.968$ & $0.999$ \\ 
TBASS & $0.022$ & $0.103$ & $0.492$ & $0.846$ & $0.926$ & $0.952$ & $0.980$ \\ 
BASS & $0.026$ & $0.122$ & $0.511$ & $0.842$ & $0.914$ & $0.950$ & $0.977$ \\ 
BART & $0.036$ & $0.147$ & $0.526$ & $0.830$ & $0.906$ & $0.940$ & $0.982$ \\ 
laGP & $0.083$ & $0.477$ & $0.176$ & $0.298$ & $0.378$ & $0.447$ & $0.580$ \\ 
\hline \\[-1.8ex] 
\end{tabular} 
\end{table} 

\begin{table}[!htbp] \centering 
  \caption{$\nu = 20, \kappa=-1.00$} 
  \label{tab:smpiston4} 
\begin{tabular}{@{\extracolsep{5pt}} cccccccc} 
\\[-1.8ex]\hline 
\hline \\[-1.8ex] 
 & RMSE\textasteriskcentered  & KS-dist & Cov 0.5 & Cov 0.8 & Cov 0.9 & Cov 0.95 & Cov 0.99 \\ 
\hline \\[-1.8ex] 
GBASS & $0.026$ & $0.099$ & $0.423$ & $0.775$ & $0.905$ & $0.969$ & $0.999$ \\ 
TBASS & $0.022$ & $0.094$ & $0.477$ & $0.809$ & $0.926$ & $0.964$ & $0.984$ \\ 
BASS & $0.025$ & $0.111$ & $0.475$ & $0.818$ & $0.925$ & $0.959$ & $0.984$ \\ 
BART & $0.032$ & $0.130$ & $0.483$ & $0.824$ & $0.908$ & $0.957$ & $0.980$ \\ 
laGP & $0.076$ & $0.461$ & $0.157$ & $0.301$ & $0.382$ & $0.453$ & $0.546$ \\ 
\hline \\[-1.8ex] 
\end{tabular} 
\end{table} 

\begin{table}[!htbp] \centering 
  \caption{$\nu = 40, \kappa=0.86$} 
  \label{tab:smpiston5} 
\begin{tabular}{@{\extracolsep{5pt}} cccccccc} 
\\[-1.8ex]\hline 
\hline \\[-1.8ex] 
 & RMSE\textasteriskcentered  & KS-dist & Cov 0.5 & Cov 0.8 & Cov 0.9 & Cov 0.95 & Cov 0.99 \\ 
\hline \\[-1.8ex] 
GBASS & $0.027$ & $0.094$ & $0.447$ & $0.787$ & $0.918$ & $0.981$ & $1$ \\ 
TBASS & $0.023$ & $0.086$ & $0.494$ & $0.805$ & $0.899$ & $0.954$ & $0.995$ \\ 
BASS & $0.026$ & $0.101$ & $0.517$ & $0.803$ & $0.908$ & $0.941$ & $0.993$ \\ 
BART & $0.033$ & $0.112$ & $0.514$ & $0.797$ & $0.915$ & $0.955$ & $0.993$ \\ 
laGP & $0.080$ & $0.463$ & $0.158$ & $0.294$ & $0.375$ & $0.455$ & $0.560$ \\ 
\hline \\[-1.8ex] 
\end{tabular} 
\end{table} 

\begin{table}[!htbp] \centering 
  \caption{$\nu = 100, \kappa=0.39$} 
  \label{tab:smpiston6} 
\begin{tabular}{@{\extracolsep{5pt}} cccccccc} 
\\[-1.8ex]\hline 
\hline \\[-1.8ex] 
 & RMSE\textasteriskcentered  & KS-dist & Cov 0.5 & Cov 0.8 & Cov 0.9 & Cov 0.95 & Cov 0.99 \\ 
\hline \\[-1.8ex] 
GBASS & $0.021$ & $0.086$ & $0.505$ & $0.813$ & $0.917$ & $0.956$ & $0.984$ \\ 
TBASS & $0.019$ & $0.127$ & $0.676$ & $0.916$ & $0.961$ & $0.980$ & $0.996$ \\ 
BASS & $0.022$ & $0.126$ & $0.612$ & $0.871$ & $0.934$ & $0.957$ & $0.977$ \\ 
BART & $0.040$ & $0.187$ & $0.585$ & $0.842$ & $0.908$ & $0.943$ & $0.971$ \\ 
laGP & $0.078$ & $0.477$ & $0.150$ & $0.314$ & $0.408$ & $0.471$ & $0.588$ \\ 
\hline \\[-1.8ex] 
\end{tabular} 
\end{table}


\end{document}